\documentclass[12pt]{article}
\usepackage{graphicx,epsfig,epsf}

\pdfoutput=1

\usepackage{graphicx}
\usepackage{scicite}
\usepackage{times}
\usepackage{rotating}
\usepackage{hyperref}
\usepackage{breakurl} 

\def\ltsima{$\; \buildrel < \over \sim \;$}
\def\simlt{\lower.5ex\hbox{\ltsima}} 
\def\gtsima{$\; \buildrel > \over \sim \;$}
\def\simgt{\lower.5ex\hbox{\gtsima}} 

\def\deg{\hbox{$^\circ$}}

\def\phflux{photons cm$^{-2}$ s$^{-1}$}

\def\Ms{$M_\odot$}

\def\r95{$r_{\rm 95}$}
\def\p0{$\pi^{\rm 0}$}
\def\t0{$t_{\rm s}$}

\def\gray{$\gamma$-ray}
\def\grays{$\gamma$-rays}
\def\Fermi{Fermi}
\def\etal{{\it et al.}}

\def\ncyg{{\rm V407 Cyg}}
\def\nsco{{\rm V1324 Sco}}
\def\nmon{{\rm V959 Mon}}
\def\ndel{{\rm V339 Del}}

\def\aap{{\it Astron.~Astrophys.}}
\def\aaps{{\it Astron.~Astrophys.~Suppl.}}
\def\actaa{{\it Acta Astron.}}

\def\aph{{\it Astropart.~Phys.}}
\def\apj{{\it Astrophys.~J.}}
\def\apjl{{\it Astrophys.~J.~Lett.}}
\def\apjs{{\it Astrophys.~J.~Suppl.~Ser.}}

\def\atel{{\it The Astronomer's Telegram}}
\def\basi{{\it Bull.~Astron.~Soc.~India}}
\def\mnras{{\it Mon.~Not.~R.~Astron.~Soc.}}
\def\nat{{\it Nature}}

\def\rmp{{\it Rev.~Mod.~Phys.}}
\def\sci{{\it Science}}

\newenvironment{sciabstract}{ \begin{quote} \bf} {\end{quote}}

\newcounter{lastnote}

\title{Fermi Establishes Classical Novae as a Distinct Class of Gamma-Ray Sources}

\author{The Fermi-LAT Collaboration\footnote{All authors with their 
affiliations appear at the end of this paper.
$\dagger$To whom correspondence should be addressed.
E-mail: Teddy.Cheung@nrl.navy.mil (C.~C.~Cheung);
E-mail: Pierre.Jean@irap.omp.eu (P.~Jean);
E-mail: shore@df.unipi.it (S.~N.~Shore)}\,\,$\dagger$}

\date{{\it Science}, submitted March 26; accepted June 20, 2014}

\begin{document}

\baselineskip15pt

\maketitle 

\vspace{-0.4in}

\begin{sciabstract} A classical nova results from runaway thermonuclear 
explosions on the surface of a white dwarf that accretes matter from a low-mass 
main-sequence stellar companion. In 2012 and 2013, three novae were detected in 
$\gamma$ rays and stood in contrast to the first \gray\ detected nova V407 Cygni 
2010, which belongs to a rare class of symbiotic binary systems. Despite 
likely differences in the compositions and masses of their white dwarf 
progenitors, the three classical novae are similarly characterized as soft 
spectrum transient \gray\ sources detected over 2$-$3 week durations. The 
$\gamma$-ray detections point to unexpected high-energy particle acceleration 
processes linked to the mass ejection from thermonuclear explosions in an 
unanticipated class of Galactic \gray\ sources. \end{sciabstract}

The \Fermi-LAT [Large Area Telescope; \cite{atw09}], launched in 2008, 
continuously scans the sky in $\gamma$ rays, thus enabling searches for 
transient sources. When a nova explodes in a symbiotic binary system, the ejecta 
from the white dwarf surface expand within the circumstellar wind of the red 
giant companion and high-energy particles can be accelerated in a blast wave 
driven in the high-density environment \cite{tat07} so that variable \gray\ 
emission can be produced, as was detected at $>$100 MeV energies by the LAT in 
V407 Cygni 2010 (\ncyg) \cite{v407}. In a classical nova, by contrast, the ejecta quickly expand 
beyond the confines of the compact binary into a much lower density environment. 
High-energy particle acceleration could therefore be related to a bow shock 
driven by the ejecta in the interstellar medium, or to turbulence and eventually 
weaker internal shocks formed in the inhomogeneous ejecta itself. The 
contribution of such expanding nova shells to cosmic-ray acceleration had been 
considered \cite{gin64}, but no predictions have so far been made for $>$100 MeV 
\grays. The classical novae (or simply ``novae" where appropriate) detected by 
the LAT with 12$-$20$\sigma$ significances (Table~1, Fig.~1) -- V959 Monocerotis 
2012 (\nmon), V1324 Scorpii 2012 (\nsco), and V339 Delphini 2013 (\ndel) -- were 
unanticipated. These observed $\gamma$ rays have higher energies than nuclear 
line emission by radioactive decay at $\sim$MeV energies that remain undetected 
in individual novae \cite{her08} and $\simlt$0.1 MeV emission detected in 
isolated cases \cite{tak09}.

\nmon\ was detected as a transient $\gamma$-ray source in June 2012 by the LAT 
while close ($\sim 20\deg$ separation) to the Sun \cite{cheungmon1} and then 
optically in August \cite{fuj12}. Ultraviolet spectroscopy revealed an 
oxygen-neon nova \cite{sho13}, recognized as the class with the most massive 
white dwarfs ($\simgt$1.1 \Ms) with massive ($\simgt$8 \Ms) progenitors [e.g., 
\cite{sta86}]. The expected peak visual magnitude of $\sim$5 would have been 
observable with the naked eye $\sim$50 days earlier, when the \gray\ transient 
was detected \cite{sho13}. \ndel\ \cite{ita13} was detected in August 2013 in a 
LAT pointed observation triggered by its high optical brightness [4.3 mag at 
peak; \cite{hay13,SOM}]. Optical spectra of \ndel\ suggest a carbon-oxygen nova 
\cite{sho13del1}, which are more common than the oxygen-neon types, with less 
massive white dwarfs evolved from $\simlt$8 \Ms\ main-sequence progenitors. 
Optical brightening of \nsco\ was detected in May 2012 \cite{wag12} and found in 
LAT \gray\ data from June \cite{cheungsco}. Although the type for \nsco\ is 
currently unclear, its optical spectroscopic evolution at early times 
\cite{wag12} did not resemble oxygen-neon novae at similar stages. We take this 
to indicate it is likely a carbon-oxygen type.

The LAT data \cite{SOM} for the three classical novae are discussed together 
with an updated analysis of the originally detected symbiotic nova \ncyg\ 
\cite{v407}. The \gray\ light curves of all four systems (Fig.~2) are similar, 
with 2$-$3 day long peaks occurring 3$-$5 days after the initial LAT detections. 
The observed optical peak preceded the \gray\ peak by $\sim$2 days in \nsco\ 
\cite{hil12,SOM} and $\sim$6 days in \ndel\ \cite{hay13,SOM}. Because the early 
optical light variations of the ejecta in novae are driven by line opacity 
changes in the ultraviolet during the expansion, the rise to peak optical 
brightness coincides with the maximum flux redistribution toward lower energies 
as the optically thick surface moves outward [see \cite{sho12}]. The initial 
lack of detected $\gamma$ rays could be because the ejecta are opaque and any 
$>$100 MeV emission produced are absorbed via photon-atom
interactions, with $\gamma$ rays appearing only later when the density drops and 
the ejecta become transparent. The three novae were detected 
in $\gamma$ rays during a time of high X-ray and ultraviolet/optical opacity. 
Coincidentally, the few days' delay of the \gray\ peak relative to the optical 
peak was also observed in \ncyg, but this may instead signal interactions with 
its red giant companion (below).

In compact classical nova binaries, typical companion separations are $a 
\sim$$10^{11}$ cm [$\sim$100$\times$ larger in symbiotic systems; \cite{mun90}] 
and expansion velocities $v_{\rm ej}$ at early times are many 100's to 
$\simgt 1000~{\rm km\,s^{-1}}$. Thus the ejecta reach the companion on a 
timescale $t=1000\,(a/10^{11}\,{\rm cm})$ $(v_{\rm ej}/1000\,{\rm 
km\,s^{-1}})^{-1}$ s (i.e., of order an hour or less). Modeling of the optical 
line profiles indicates that the spatial distribution of the ejected gas is 
bipolar rather than spherical in all cases, with greater extension 
perpendicular to the orbital plane in \nmon\ \cite{sho13,rib13,her92}. Also, 
narrow absorption and emission line structures seen in optical and ultraviolet 
line profiles later in the expansion may be evidence of hydrodynamical 
instabilities and multiple ejections that may lead to the formation of strong 
turbulence and internal shocks within the ejecta after the ignition of the 
thermonuclear runaway \cite{cas11}. A clue to the physical process that causes 
the \gray\ emission mechanism may be the similarity of the high-energy spectral 
characteristics of \nsco, \nmon, and \ndel. Their $>$100 MeV spectra are all 
soft, and can be fit with single power laws (the spectrum $N$($E$) $\propto 
E^{-\Gamma}$ with the number $N$ of photons with energy $E$) with photon 
indices $\Gamma=$ 2.1$-$2.3, or exponentially cutoff 
power laws (the spectrum $N$($E$) $\propto E^{-s}e^{-E/E_{\rm c}}$, where 
$E_{\rm c}$ is the cutoff energy) -- see \cite{SOM}, Table~S1 and Fig.~S1. The 
exponentially cutoff power law fits to the LAT data were preferred over the 
power law fits at the 3.8$\sigma$ and $3.4\sigma$ level for \nmon\ and \ndel, 
respectively, but provided an insignificant improvement (2.0$\sigma$) for \nsco. 
Considering the uncertainties in the spectral fits, the three novae are 
similarly characterized by slopes $s=$ 1.7$-$1.8, $E_{\rm c}\sim$ 1$-$4 GeV, and 
observed emission up to $\sim$6$-$10 GeV. The total durations of the observed 
$\gamma$ rays were also similar, being detected for 17$-$27 days at $>$2$\sigma$ 
statistical significances in daily bins (Fig.~2, Table~1). Because the 
LAT-observed properties are similar, it is likely that the \gray\ emission of 
these classical novae has a similar origin, involving interactions of the 
accelerated high-energy protons (hadronic scenario) or electrons (leptonic 
scenario) within the ejecta.

In the hadronic scenario, high-energy protons that interact with nuclei produce 
neutral pions (\p0), which decay into two $\gamma$ rays. For a representative 
hadronic model, we assume an exponentially cutoff power law distribution of 
protons in the form, $N_{\rm p} (p_{\rm p}) = N_{\rm p,0} \, (p_{\rm p} \, 
c)^{-s_{\rm p}} \, e^{-W_{\rm p}/E_{\rm cp}}$ (proton/GeV), where $p_{\rm p}$ 
and $W_{\rm p}$ are the momentum and the kinetic energy of protons, 
respectively, $N_{\rm p,0}$ the normalization, $s_{\rm p}$ the slope, and 
$E_{\rm cp}$ the cutoff energy. We fitted $E_{\rm cp}$ and $s_{\rm p}$ with the 
LAT spectra to obtain the best-fit \p0\ models (Fig.~3). The lower limits to the 
cutoff energies ($\sim$3$-$30 GeV) suggest proton acceleration up to near-TeV 
energies. The slopes of the best-fit models of the proton spectrum have large 
statistical uncertainties ($\sim 0.8$) but interestingly are compatible with a 
value of 2 expected in the first order Fermi acceleration process. To 
match the observed $\gamma$-ray variability timescale in such a 
process, a magnetic field $B > 10^{-3}$ Gauss is required in a strong shock with 
$v_{\rm ej} = 2000$ km s$^{-1}$ to accelerate particles to $> 1 (10)$ GeV 
energies in $\sim 0.2 (2)$ days. Formally, the updated best-fit proton spectrum 
for the symbiotic nova \ncyg\ [cf., \cite{v407}] is parameterized by $s_{\rm 
p}=1.4^{+0.3}_{-0.4}$ GeV, but slopes of 2 -- 2.2 are also viable at 
the 90$\%$ confidence level with $E_{\rm cp} = 10^{+1.0}_{-0.7}$ GeV 
[\cite{SOM}, Fig.~S3]. Lower-confidence fits were also obtained for \nmon\ and 
\ndel\ but conversely with smaller slopes and lower cutoff energies [\cite{SOM}, 
Fig.~S3]. Assuming that the \gray\ flux is due to the interactions of 
high-energy protons with the nuclei in the ejecta, the best-fit parameters allow 
us to estimate the total energy in high-energy protons of $\sim$(3$-$17)~$\times 
10^{42}$ ergs and to derive conversion efficiencies (i.e., the ratio of the 
total energy in high-energy protons to the kinetic energy of the ejecta) ranging 
from $\sim$0.1$-$3.7$\%$ for the classical novae and 6.6$\%$ for \ncyg.

In the leptonic case, accelerated electrons produce $\gamma$ rays through a 
combination of inverse Compton scattering with low-energy photons and 
bremsstrahlung with atoms in the vicinity of the nova. For a leptonic model, we 
adopted a similar functional form for the distribution of the kinetic energy of 
high-energy electrons ($W_{\rm e}$) in the form $N_{\rm e}(W_{\rm e}) = N_{\rm 
e,0} W_{\rm e}^{-s_{\rm e}} e^{-W_{\rm e} / E_{\rm ce}}$ (electron/GeV), and 
fitted the normalization $N_{\rm e,0}$, slope $s_{\rm e}$, and cutoff energy 
$E_{\rm ce}$ to the LAT data for each nova (Fig.~3). The \gray\ luminosity of 
the calculated bremsstrahlung emission is $<$20$\%$ of the total \gray\ 
luminosity for all the novae \cite{SOM}. The best-fit parameters of the 
high-energy electron spectra for the three classical novae are similar within 
their confidence regions \cite{SOM}, with $E_{\rm ce}$ constrained to lie 
between 2 and 30 GeV, and poorly constrained slopes. These models are 
statistically indistinguishable from the \p0\ model. As in the hadronic model, 
the spectral parameters of the classical novae differ from those for \ncyg\ 
(mainly due to the lowest-energy $\sim$ 200$-$300 MeV bin detected in its LAT 
spectrum) where the best-fit slope is negative (i.e., a positive index of the 
power law) and $E_{\rm ce}$ = $1.78\pm0.05$ GeV. The best-fit parameters for the 
leptonic scenario, where high-energy electrons interact primarily with the 
photons emitted by the nova photosphere \cite{mar13}, lead to total energies of 
$\sim$(6$-$13)~$\times 10^{41}$ ergs in high-energy electrons and conversion 
efficiencies of $\sim0.1 - 0.3\%$ for the classical novae and 0.6$\%$ for the 
symbiotic system.

Detection of classical novae in $\gamma$ rays was deemed unlikely in the past 
\cite{v407}. The only nova previously detected in $\gamma$ rays, the 
aforementioned \ncyg, was a rare symbiotic and likely recurrent [only 10 
recurrent novae are known, of which 4 are symbiotic types; \cite{sch10}]. In the 
symbiotic novae, conditions are conducive for high-energy particle acceleration 
as the portion of the ejecta moving into the wind in the direction of the dense 
medium provided by the red giant companion decelerates within a few days. The 
\grays\ peak early, when the efficiency for hadron and lepton acceleration is 
presumably favorable, with the red giant wind playing a key role in the \gray\ 
production [see \cite{tat07,mar13}]. In contrast, the main-sequence star 
companions in the classical novae do not provide similarly dense target 
material, hence it is likely that other dissipative processes are involved in 
particle acceleration and generation of the observed $\gamma$ rays.

Because the \gray\ properties of the novae detected so far by the \Fermi-LAT 
appear similar to one another, and their underlying properties are unremarkable, 
it appears all novae can be considered to be candidate $\gamma$-ray emitters. 
Their detection by the LAT may imply close proximity and that other optical 
novae not yet detected with the LAT [e.g., \cite{cheungfermi}] are more distant 
and have fainter optical peaks [without considering extinction uncertainties 
\cite{mukai}]. Indeed, all the LAT-detected novae have estimated distances of 
$\simlt$4$-$5 kpc (Table~1). Despite systematic uncertainties in the adopted 
distances, it is interesting that the inferred mean \gray\ luminosities and 
total emitted energies of the novae span a small range $\sim$(3$-$4)$\times 
10^{35}$ ergs s$^{-1}$ and $\sim$(6$-$7)$\times 10^{41}$ ergs, respectively, 
except for the $\sim$2$\times$ greater values for \nsco\ whose distance is 
highly uncertain.

The rate of novae in the Milky Way is highly uncertain, but considering a 
plausible range of $\sim$20$-$50 per year \cite{sha97} and reasonable spatial 
distributions in the Galactic bulge and disk \cite{jea00}, our estimate is 1$-$4 
per year at $\simlt$4$-$5 kpc distances. The \gray\ detection rate of novae 
averages roughly to once per year over the timespan of these observations 
($\sim$5 years), consistent with the lower end of this extrapolation.

Although the \gray\ properties of the LAT-detected novae are similar, we 
emphasize the small and subtle differences that imply different emission 
mechanisms, e.g., the spectral shape of \ncyg\ compared to the three classical 
novae as well as the apparent higher energy extension of the \nsco\ spectrum. 
Among the classical novae detected so far, they also appear different optically. 
The \gray\ emission mechanism and high-energy particle acceleration processes 
associated with the novae could depend on the particular system properties that 
remain to be investigated, such as the white dwarf mass, which determines the 
explosion energetics (ejected mass, expansion velocity), and the mass transfer 
dictated by the companion mass and separation.


\noindent {\bf Acknowledgments:} The \Fermi-LAT Collaboration acknowledges 
support for LAT development, operation and data analysis from NASA and DOE 
(United States), CEA/Irfu and IN2P3/CNRS (France), ASI and INFN (Italy), MEXT, 
KEK, and JAXA (Japan), and the K.A.~Wallenberg Foundation, the Swedish Research 
Council and the National Space Board (Sweden). Science analysis support in the 
operations phase from INAF (Italy) and CNES (France) is also gratefully 
acknowledged.
C.C.C.~was supported at NRL by a Karles' Fellowship and by NASA through 
DPR S-15633-Y and Guest Investigator programs 11-FERMI11-0030 and 
12-FERMI12-0026.
S.S.~was supported by NASA and NSF grants to ASU.
The \Fermi-LAT data reported in this paper are available from 
\url{http://fermi.gsfc.nasa.gov/ssc/data/access/}.


\begin{table*}
\small
\footnotesize
  \begin{center}
    \tabcolsep 1.75pt
\begin{tabular}{lcccc}
\hline\hline
\multicolumn{1}{l}{Nova} &
\multicolumn{1}{c}{V407 Cyg 2010} &
\multicolumn{1}{c}{V1324 Sco 2012} &
\multicolumn{1}{c}{V959 Mon 2012} &
\multicolumn{1}{c}{V339 Del 2013} \\
\hline
\hline
Distance (kpc)      & 2.7 & 4.5 & 3.6 & 4.2 \\
Peak magnitude      & 6.9 & 10.0 & 5*  & 4.3 \\
Peak date           & 2010 Mar 10.80 & 2012 Jun 19.96 &  ... & 2013 Aug 16.50 \\
Optical RA, Decl.\  & 315.5409\deg, +45.7758\deg\ & 267.7246\deg, --32.6224\deg\ &  99.9108\deg, +5.8980\deg\ & 305.8792\deg, +20.7681\deg\ \\
Optical $l$, $b$    & 86.9826\deg, --0.4820\deg\  & 357.4255\deg, --2.8723\deg\  & 206.3406\deg, +0.0754\deg\ & 62.2003\deg, --9.4234\deg\ \\
\hline
LAT RA, Decl.\      & 315.57\deg, +45.75\deg\ & 267.72\deg, --32.69\deg\ & 99.98\deg, +5.86\deg\ & 305.91\deg, +20.78\deg\ \\
Optical-LAT offset  & 0.03\deg\ & 0.07\deg\ & 0.08\deg\ & 0.03\deg\ \\
LAT error radius (95$\%$) & 0.08\deg\ & 0.09\deg\ & 0.18\deg\ & 0.12\deg\ \\
\t0\ (date)         & 2010 Mar 10 & 2012 Jun 15 &  2012 Jun 19 & 2013 Aug 16 \\
\t0\ (MJD)          & 55265 & 56093 & 56097 & 56520 \\
Duration (days)     & 22 & 17 & 22 & 27 \\
$L_{\gamma}$ ($10^{35}$ erg s$^{-1}$) & 3.2 & 8.6 & 3.7 & 2.6 \\
Total energy ($10^{41}$ erg)          & 6.1 & 13 & 7.1 & 6.0 \\
\hline
\hline
\end{tabular}
\normalsize
\label{table-summary}
  \end{center}
{\bf Table~1 -- Summary of the Four Novae.} 
Tabulated are optical peak magnitudes and adopted 
distances from \cite{mun90} for \ncyg, estimate of $\sim$4$-$5 kpc \nsco\ based 
on the maximum magnitude rate of decline relation \cite{hil12} notwithstanding 
the large uncertainties in this method \cite{kas11}, \cite{sho13} for \nmon\ 
(scaled from V1974 Cyg 1992), and \cite{sho13del2} for \ndel\ (scaled from OS 
Andromedae 1986), and observed dates of the optical peaks
(unfiltered from \cite{v407}, $V$-band, adopted, and visual magnitudes, respectively).
Positions in J2000.0 equinox (right ascension, RA; declination, Decl.), Galactic 
longitude ($l$) and latitude ($b$), 95$\%$ confidence localization error radius, 
and offset between the LAT and optical positions in units of degrees. Adopted 
start dates \t0\ \cite{SOM} are given in Gregorian Dates and Modified Julian Days 
(MJD). The \gray\ luminosities $L_{\gamma}$ and total emitted energies were 
estimated with the average fluxes from the power law fits of the $>$100 MeV LAT 
spectra integrated up to 10 GeV and durations from \t0\ up to the last 
$>2\sigma$ daily bin LAT detection. For \ndel, the $\gamma$ rays were 
detected for 25 days in 1-day bins (Fig.~2), but there was a hint of a detection 
two days earlier on the day of the optical peak in 0.5-day binned data \cite{SOM},
leading to a 27 day duration.\\
{\rm *}Note that for \nmon\ the optical peak magnitude of 9.4 (unfiltered) was 
observed $\sim$50 days after the initial \gray\ detection, and we adopted an 
inferred peak of 5 magnitude \cite{sho13}.
\end{table*}


\begin{figure}[htbp]
  \begin{center}
    \includegraphics[width=7.75cm]{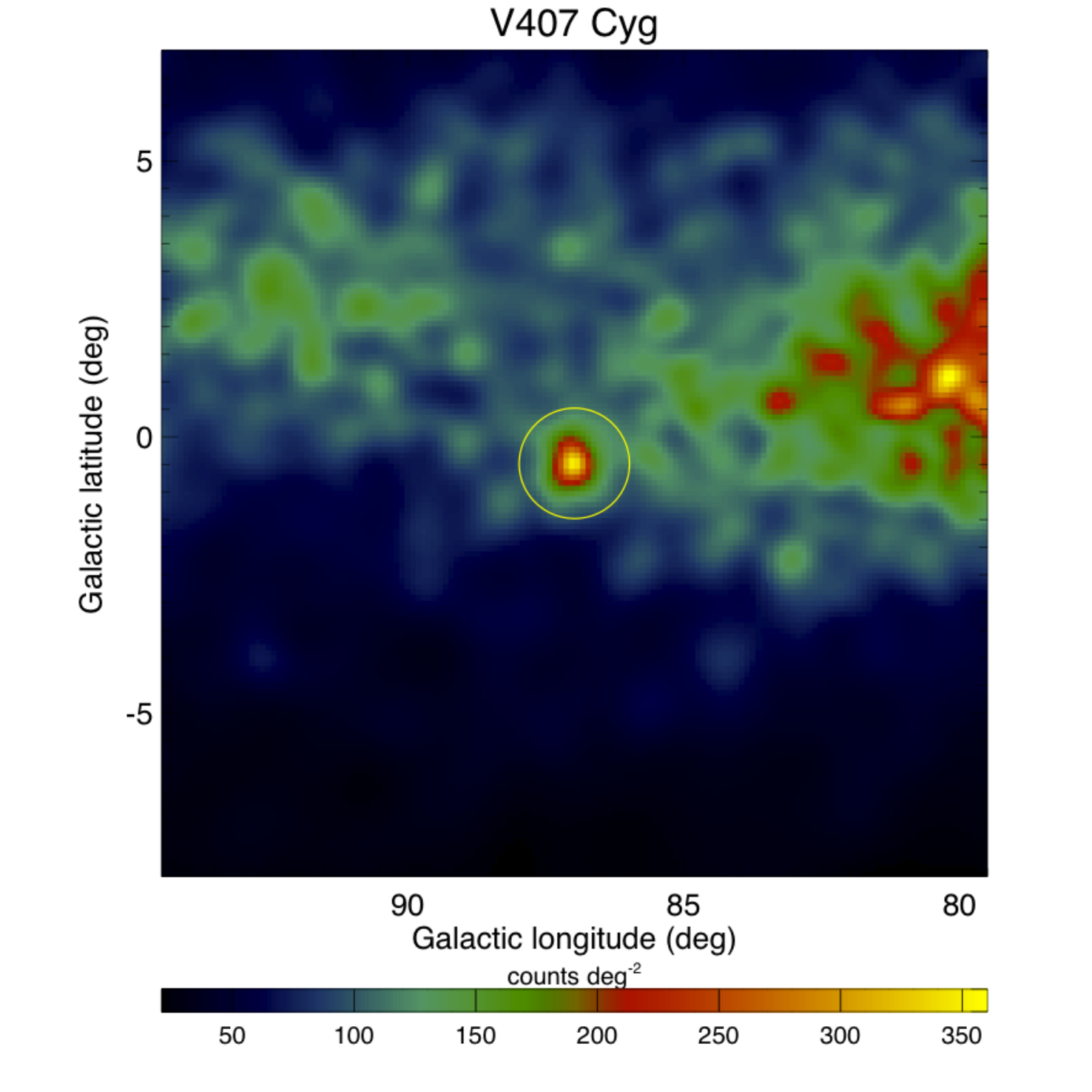}
    \includegraphics[width=7.75cm]{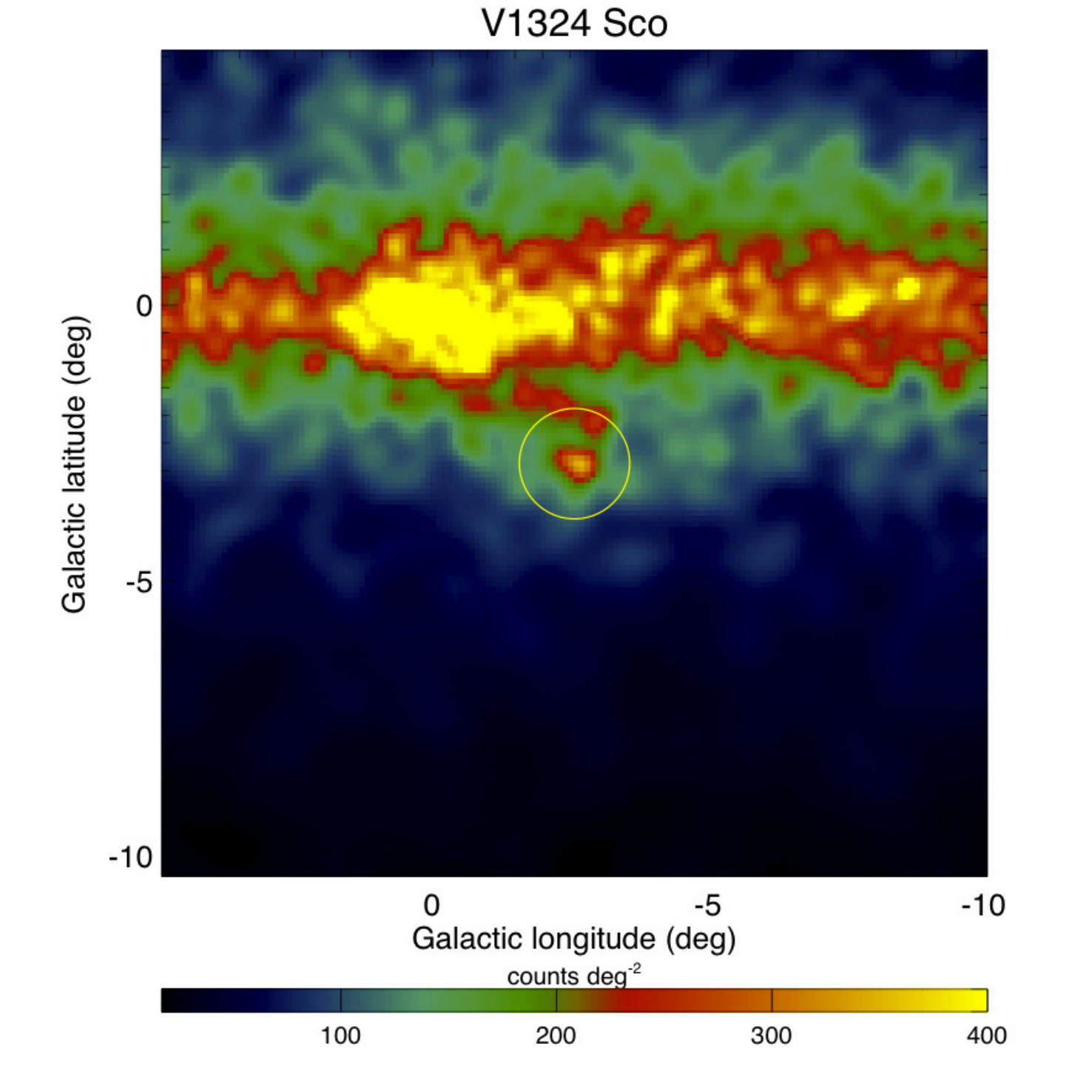}
    \includegraphics[width=7.75cm]{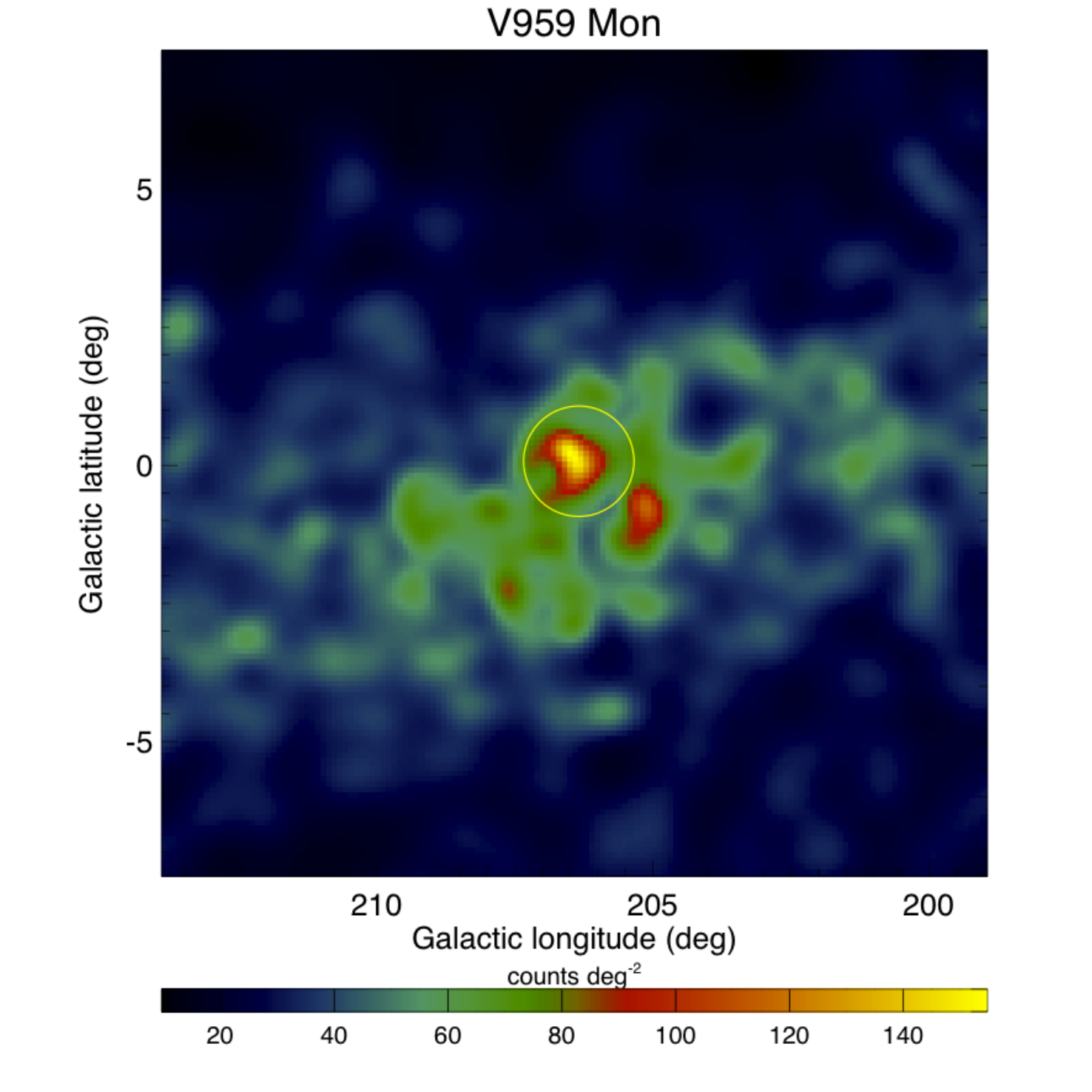}
    \includegraphics[width=7.75cm]{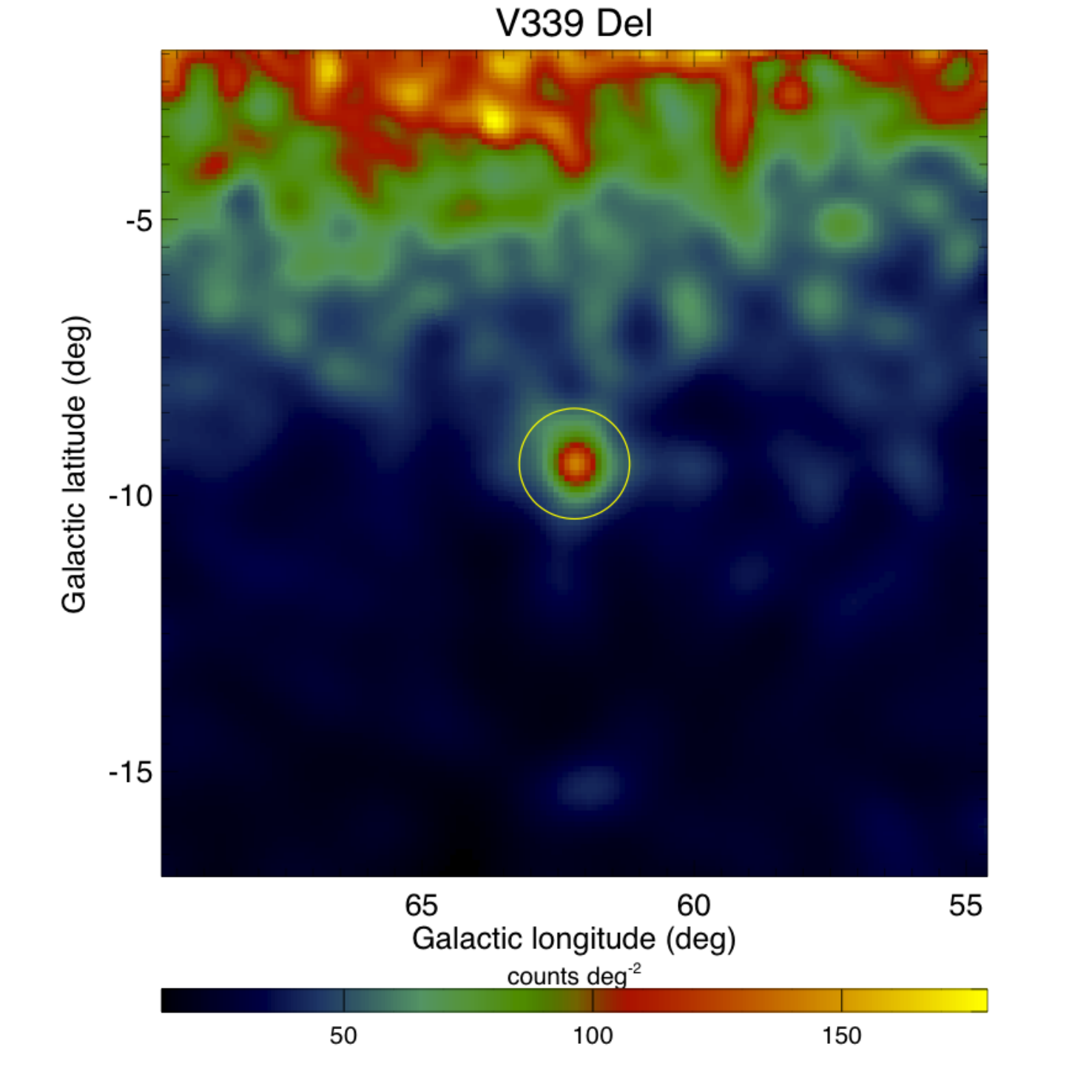}
  \end{center} {\bf Fig.~1.} \Fermi-LAT $>$100 MeV \gray\ counts maps of the 
four novae in Galactic coordinates centered on the optical positions over the 
full 17$-$27 day durations. The maps used $0.1\deg \times 0.1\deg$ pixels and 
were adaptively smoothed with a minimum number of 25$-$50 counts per kernel. Each 
nova (located at the centers of the yellow circles with 1\deg\ radius which is 
the approximate LAT $95\%$ containment at 1 GeV) is observed near the bright 
diffuse \gray\ emission in the Galactic plane, with \nmon\ in particular 
observed directly through the plane (0\deg\ latitude). 
\end{figure}

\begin{figure}[htbp]
  \begin{center}
    \includegraphics[width=16.5cm]{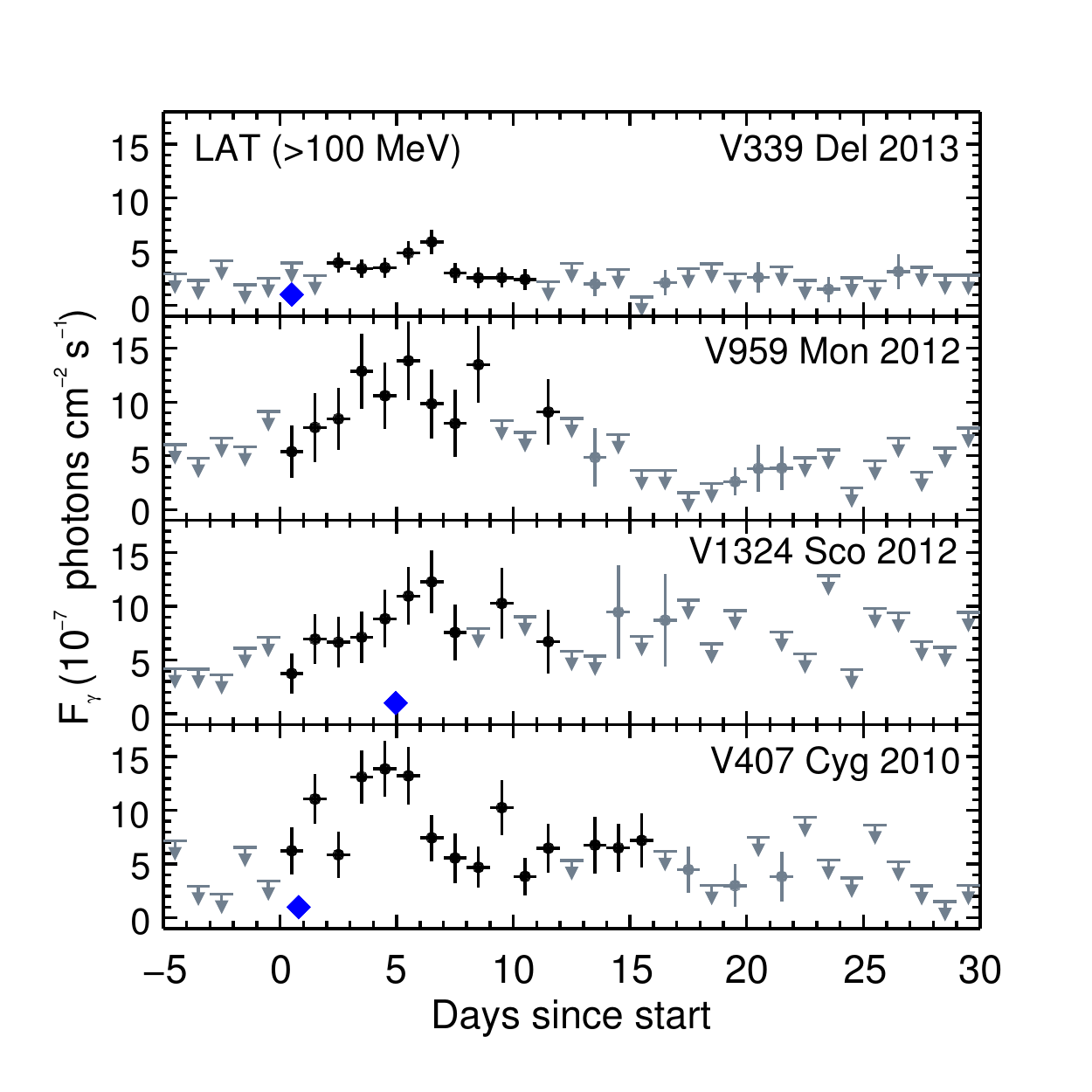}
  \end{center} {\bf Fig.~2.} \Fermi-LAT 1-day binned light curves of the four 
\gray\ detected novae. Vertical bars indicate 1$\sigma$ uncertainties for data 
points with $>3\sigma$ (black) and $2$-$3\sigma$ (gray) significances, 
otherwise, 2$\sigma$ upper limits are indicated with gray arrows. Start times 
\t0\ (from top to bottom panels) of 2013 August 16, 2012 June 19, 2012 
June 15, and 2010 March 10 were defined as the day of the first \gray\ detection.
In \ndel, there was a $2.4\sigma$ detection in 0.5-day binned data 
beginning August 16.5 \cite{SOM}, the epoch of the optical peak (blue 
diamond in each panel).
\end{figure}

\begin{figure}[htbp]
  \begin{center}
    \includegraphics[width=7.75cm,angle=-90]{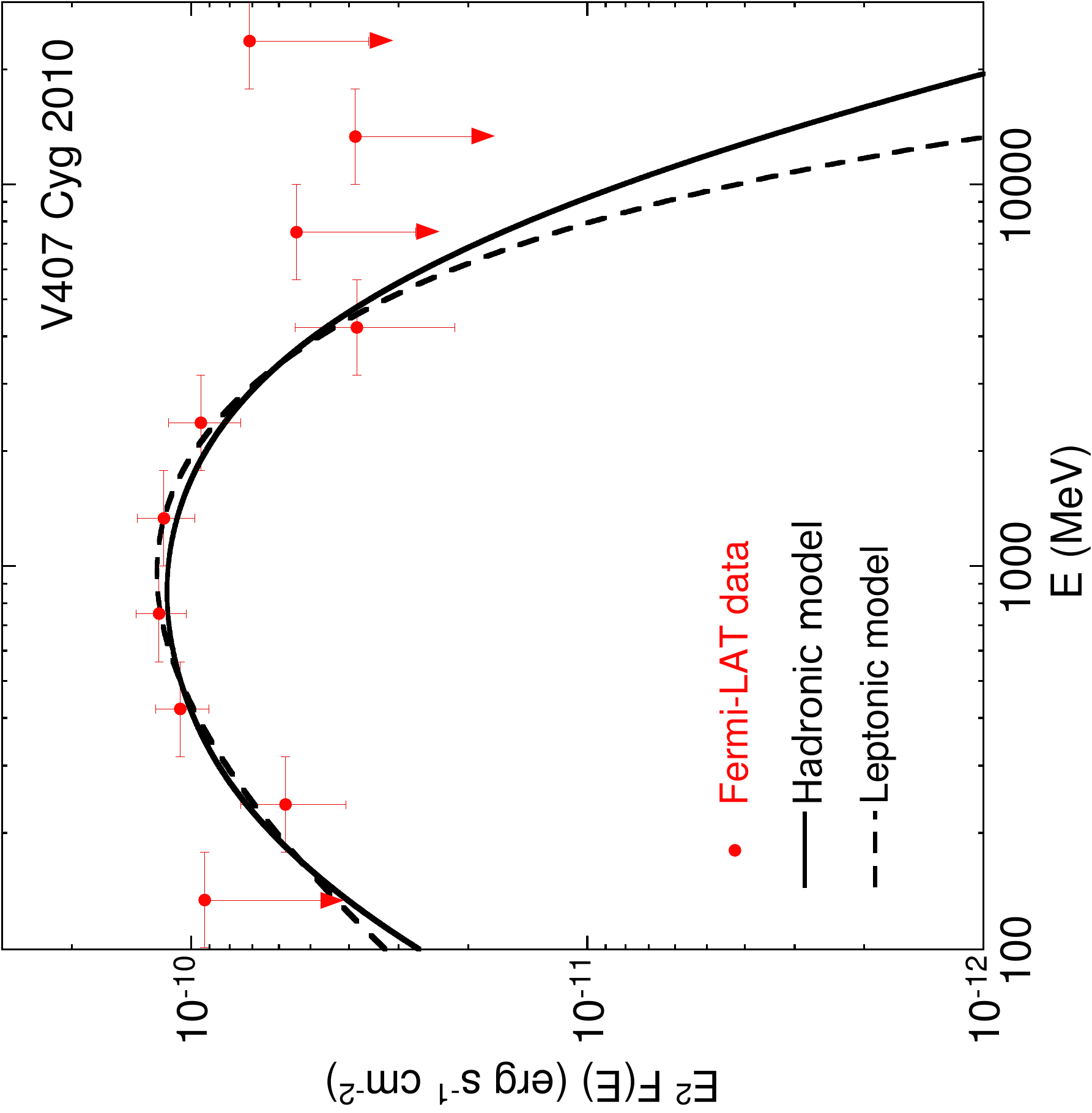}
    \includegraphics[width=7.75cm,angle=-90]{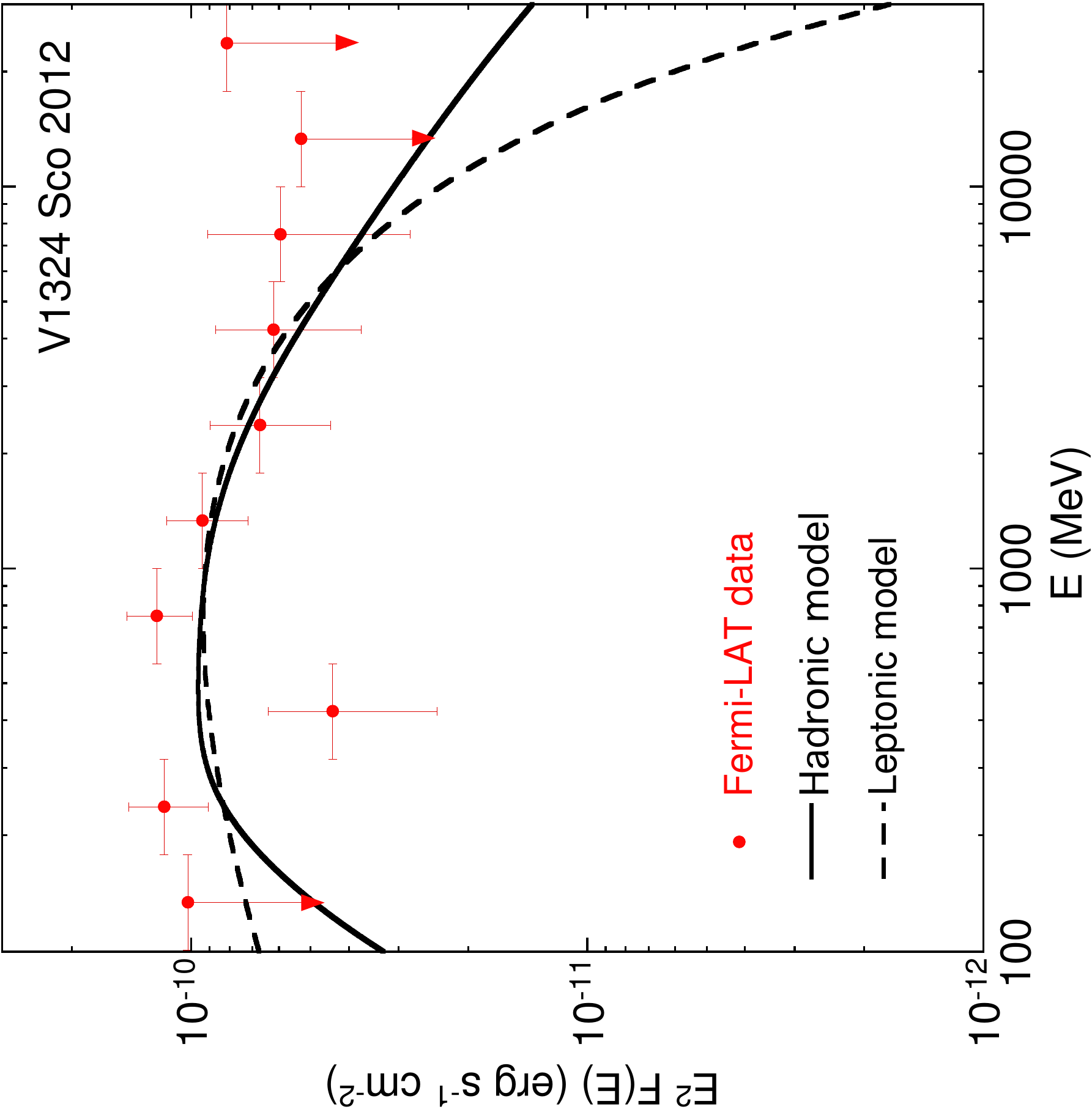}
    \includegraphics[width=7.75cm,angle=-90]{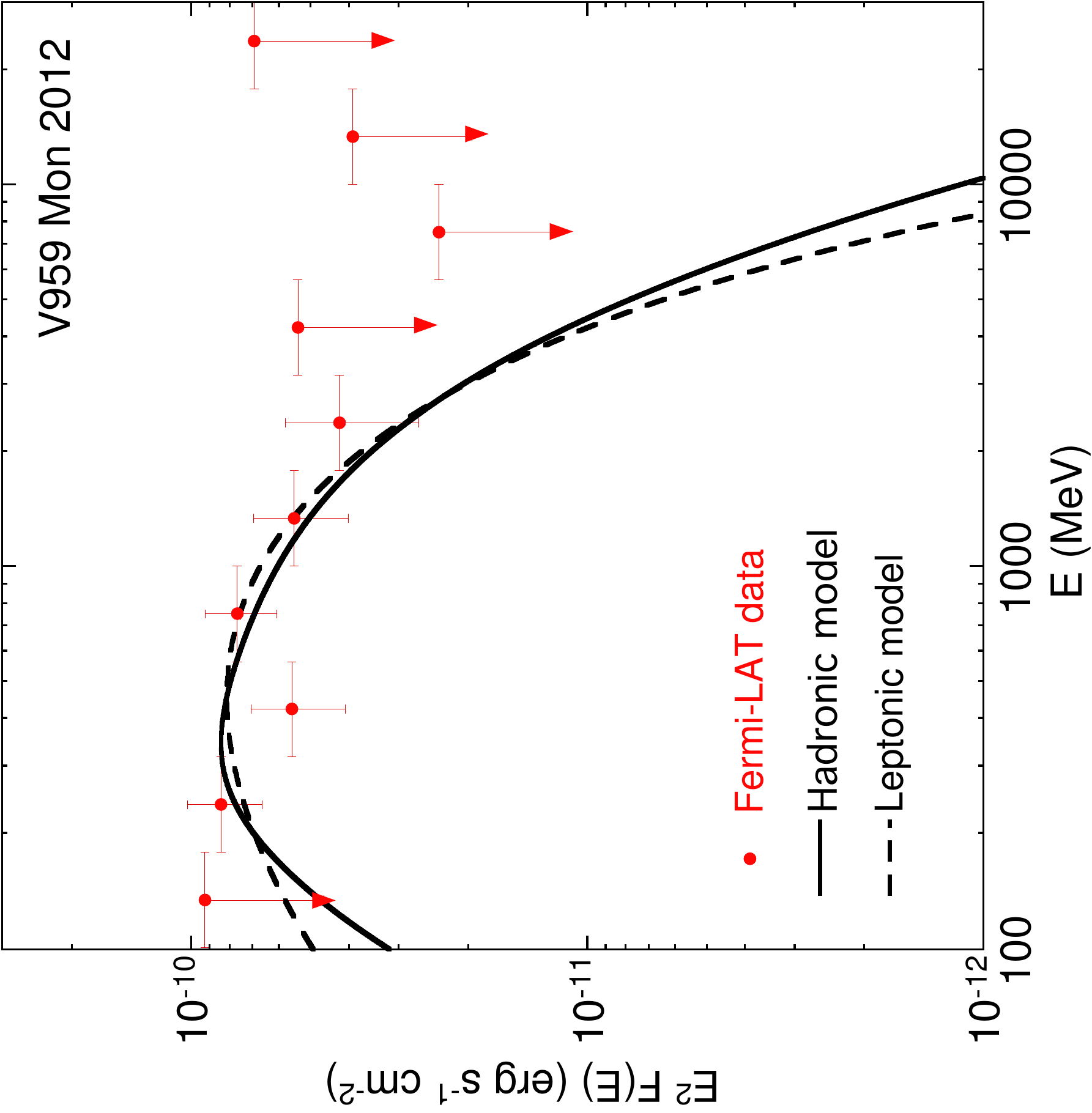}
    \includegraphics[width=7.75cm,angle=-90]{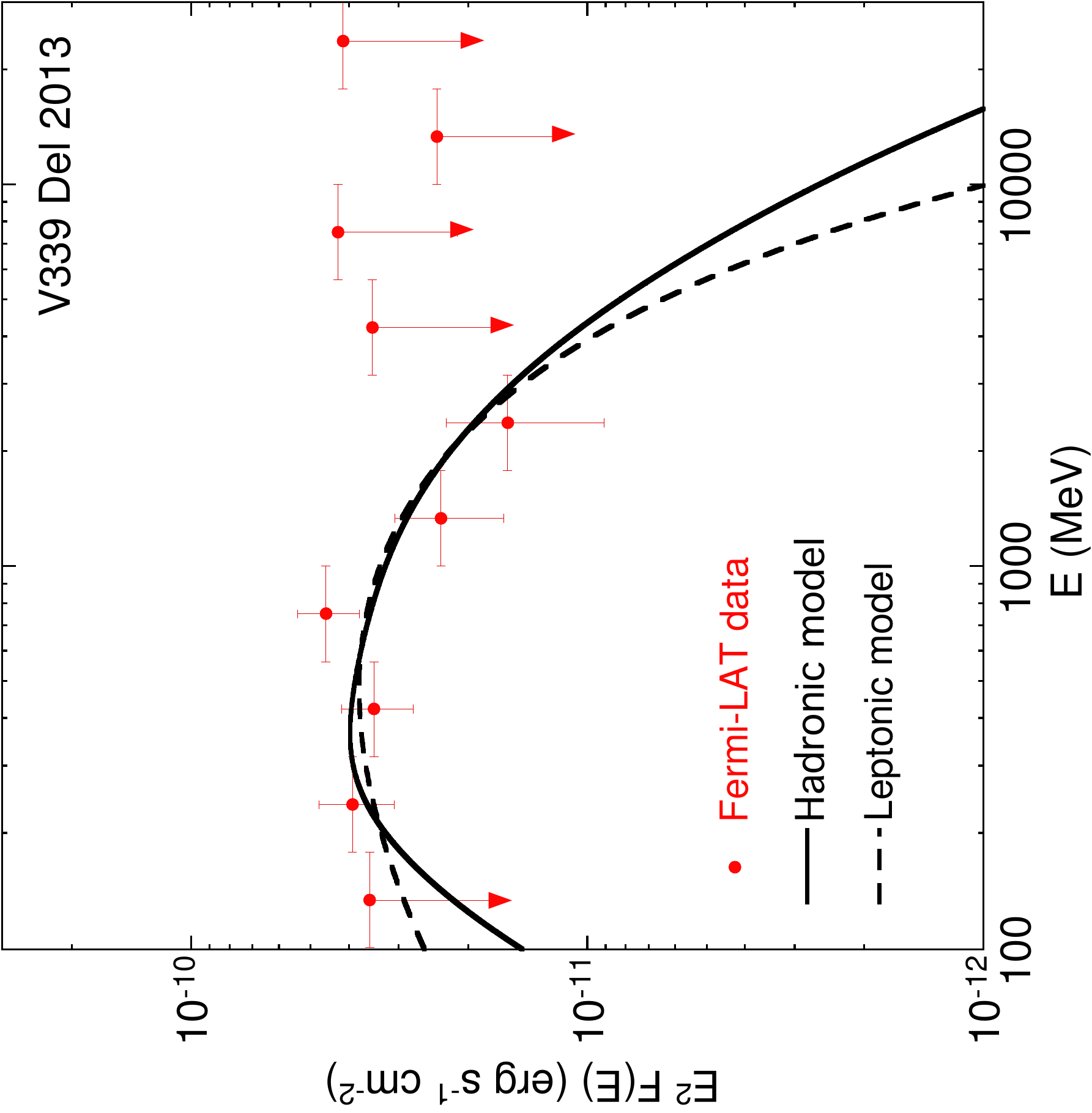}
  \end{center} {\bf Fig.~3.} \Fermi-LAT $>$100 MeV average \gray\ spectra of the 
four novae over the full 17$-$27 day durations.  Vertical bars indicate 
1$\sigma$ uncertainties for data points with significances $>2\sigma$; 
otherwise, arrows indicate 2$\sigma$ limits. The best-fit hadronic and leptonic 
model curves are overlaid. 
\end{figure}

\clearpage
\newpage
\setcounter{page}{1}

\begin{center}{\Large Supplementary Materials for}\end{center}

\begin{center}{\large Fermi Establishes Classical Novae as a Distinct Class of Gamma-Ray Sources}\end{center}

\begin{center}The Fermi-LAT Collaboration\footnote{All authors with their
affiliations appear at the end of this paper.}\end{center}

\begin{center}
$\dagger$correspondence to: 
Teddy.Cheung@nrl.navy.mil (C.~C.~Cheung),
Pierre.Jean@irap.omp.eu (P.~Jean),
shore@df.unipi.it (S.~N.~Shore)
\end{center}

\noindent
{\bf This PDF file includes:}

Materials and Methods

Figs.~S1 to S6

Tables S1 to S4

References ({\it 31-49})

\newpage

\section*{Materials and Methods}

\subsection*{S1.~\Fermi-LAT Observations and Analysis}

The \Fermi\ Large Area Telescope (LAT) \cite{atw09} is sensitive to $\gamma$ 
rays from 20 MeV to $>$300 GeV. It features a large instantaneous field of view 
(2.4 steradian) and began operations in 2008 August nominally in survey mode 
where an all-sky image is obtained every two orbits ($\sim$3 hrs). Three of the 
novae (\ncyg, \nsco, \nmon) were detected over the course of the LAT all-sky 
survey. For \ncyg, our analysis here updates that originally published 
\cite{v407} with the \gray\ event selection cuts and associated instrument 
response functions to match those used for the new cases. \nsco\ in particular 
was found in the field of a source being monitored for other reasons [see 
\cite{hil12}] and was detected in $\gamma$ rays as early as June 15 (just four 
days earlier than \nmon). Its positional and temporal coincidence with \nsco\ 
[MOA 2012-BLG-320, \cite{wag12}] was noticed later \cite{cheungsco}. \ndel\ was 
the subject of a \Fermi\ target-of-opportunity (ToO) pointing observation 
triggered by the bright optical discovery \cite{ita13}. The ToO began 2013 
August 16.5 UT, the day of the optical peak (Section~S3). The \ndel\ ToO lasted 
for six days and resulted in $\sim$3$\times$ greater exposure with the LAT than 
would have been possible in all-sky survey mode, leading to improved statistics 
that were especially useful because of the relative faintness of the nova 
(below).

For the LAT analysis, we selected 100 MeV to 100 GeV events within regions of 
interest (ROI) of 15\deg\ radius centered on each nova optical position 
(Table~1) using the Pass 7 data and {\tt P7SOURCE$\_$V6} instrument response 
functions (IRFs)\footnote{Cross-checks with the newer Pass 7 reprocessed data 
and response functions produced results consistent within the quoted 
uncertainties.}. Data were selected with a rocking angle cut of 52\deg\ and 
maximum zenith angle of 100\deg\ in order to minimize contamination from Earth 
limb photons, using {\tt gtmktime} with the filter ($\#$3) recommended for the 
combination of sky survey and ToO 
observations\footnote{\burl{http://fermi.gsfc.nasa.gov/ssc/data/analysis/ 
documentation/Cicerone/Cicerone_Likelihood/Exposure.html}}. To model the diffuse 
\gray\ background and nearby $\gamma$ ray sources we utilized the isotropic and 
Galactic diffuse emission templates\footnote{Files {\tt iso$\_$p7v6source.txt} 
and {\tt gal$\_$2yearp7v6$\_$v0.fits}, respectively.} and included all sources 
from the 2nd \Fermi\ Gamma-ray LAT [2FGL, \cite{2fgl}] catalog within the ROI. 
The analysis used version {\tt v9r27p1} of the \Fermi\ science tools assuming a 
point source for each target.

Uncertainties shown in the plots are statistical. Two major sources of 
systematic errors on the results are the uncertainties in the LAT effective area 
and the modeling of interstellar emission, since all these novae lie close to 
the Galactic plane. The uncertainties in the effective area for the IRFs we use 
are evaluated as $10\%$ at 100 MeV, $5\%$ at 560 MeV, and $10\%$ above 10 GeV, 
linearly varying with the logarithm of energy between those values 
\cite{ack12s}. In all cases, the statistical uncertainties exceed these values.  
The uncertainty due to the interstellar emission model does not affect the light 
curves but would enlarge the uncertainties on the spectral measurements.

\subsubsection*{\underline{Light curves and Durations}}

To determine the interval and duration over which \gray\ emission was detected 
for each nova, we generated LAT light curves using a binned {\tt gtlike} 
analysis in daily bins centered at fiducial start times \t0\ set to the first 
reported LAT detection for \ncyg\ \cite{v407}, \nsco\ \cite{cheungsco}, and 
\nmon\ \cite{cheungmon1}. For \ndel, the first LAT detection in 1-day bins was 
on August 18 (see Sec.~S3), but there was indication for a marginal detection 
(see below) in 0.5 day bins when the optical peak was observed two days earlier; 
thus \t0\ was set to the latter in this case. The full interval studied for each 
nova was 106 days, i.e., \t0\ $\pm 53$ days (to gauge the effect of varying 
exposure over two precession periods of the spacecraft orbit), except for \ndel, 
for which 2013 September 30 (MJD 56565) was the last day of data used (for \t0\ 
= 56520; this is \t0\ $-61$/$+45$ days). The start times defined in this way do 
not necessarily coincide with the optical discovery date, which is in contrast 
in some cases to the convention often adopted in the nova literature. We chose 
to define the start times in this way because the \gray\ emission is often 
detected at or near the optical peaks and these seem to be the relevant time 
windows to utilize for the light curve comparisons. For \ncyg, the first daily 
bin \gray\ detection coincided with the apparent optical peak observed on the 
discovery date, although there was an uncertainty in the optical peak epoch of 
up to three days due to an observing gap. In the case of \nsco\ \cite{hil12}, 
the optical discovery date of May 22.80 UT \cite{wag12} was 23 days before the 
first LAT detection, making the former an impractical choice of a start time for 
this work. For \ndel\ \cite{hay13}, the discovery date was two days before the 
optical peak when the first LAT detection occurred in 0.5-day binned data 
(below), but a further two days before the first daily LAT detection.

For all light curves, we left the normalization of the isotropic diffuse 
spectrum as a free parameter in each time bin while the Galactic normalization 
was initially fit over the full 106-day interval, then fixed at the average 
fitted value in the shorter time-bins. All 2FGL sources within each ROI were 
included in the model using their cataloged spectral parameters \cite{2fgl}. For 
the 2FGL sources within 5\deg\ of the novae positions known to be variable 
[according to \cite{2fgl}], the flux normalizations were initially fit over each 
106-day interval. The newly fitted fluxes were assumed for the variable sources 
if significantly detected with test statistic \cite{mat96s} $TS>10$, otherwise, 
the 2FGL fluxes were used in the subsequent 1-day binned analysis. We assumed a 
power-law (PL; the spectrum $N$($E$) $\propto E^{-\Gamma}$) spectral model for 
point sources fit at the optical nova positions. The 1-day binned data were 
initially fit with free normalization and photon index $\Gamma$. We found no 
statistically significant variations in the index from the average values of the 
most significant detections with $TS>9$, $\Gamma = 2.2$ for \ncyg\ ($N=15$ bins) 
and \ndel\ ($N=9$). We similarly found no significant variations from the 
average daily values $\Gamma = 2.2$ for \nsco\ ($N=7$) and 2.4 for \nmon\ 
($N=6$) considering instead a larger $TS>15$ threshold because of the brighter 
Galactic diffuse background in these cases. We then fixed the PL spectral slopes 
and regenerated the light curves to determine durations over which to sum the 
data to fit average spectra (below). The final daily light curves presented in 
Fig.~1 assumed $\Gamma = 2.2$ (\ncyg, \nsco) and 2.3 (\nmon, \ndel), with the 
latter taken to match the values fitted below over the total intervals; \ncyg\ 
was not adjusted because of the significant spectral curvature detected.

Considering a $TS>4$ threshold\footnote{The source significance is 
$\sim\sqrt{TS}$ assuming one degree of freedom.} for detections rather than 
upper limits in the 1-day binned light curves (Fig.~2; tabulated in Table~S2), 
the resultant observed durations of detectable \gray\ emission were 2$-$3 weeks 
for all sources. More precisely, the observed range was 17$-$22 days for the 
first three novae. For \ndel, $\gamma$ rays were detected for 25 days beginning 
1.5 days after the optical peak observed on August 16.5. Note however, that the 
ToO observation commenced also on August 16.5; thus we generated a LAT light 
curve with 0.5 day bins and found $TS = 0$ for the earlier bin and $TS = 5.6$ 
with $>$100 MeV flux = $(2.5\pm1.3) \times 10^{-7}$\phflux\ for the latter half 
which benefitted from the increased exposure from the ToO. We therefore defined 
the duration for \ndel\ to be 27 days. Because the varying exposures over the 
timescales of many weeks for the LAT observations may complicate statements 
about the detectability of each nova at both early and late times, this may 
affect our measured \gray\ durations, hence estimates of the total energy 
emitted in $\gamma$ rays. As a final note, applying the same $TS>4$ threshold to 
the original LAT analysis for \ncyg\ results in a \gray\ duration of 18 days 
[\cite{v407}, Table~S1 therein], and is shorter than the 22 days we found in our 
reanalysis (due to an increase in TS of two bins from $\sim$1$-$2 to 4; 
Table~S2).

\subsubsection*{\underline{Positions}}

To obtain the most precise source positions with the LAT, we performed an 
unbinned {\tt gtlike} analysis assuming a single PL model over the \gray\ active 
durations defined in the 1-day binned analysis. The resultant formal integrated 
detection significances were 12$-$20$\sigma$; the corresponding sky counts maps 
are shown in Fig.~1. We then ran {\tt gtfindsrc} and obtained $95\%$ confidence 
error radii of $0.068\deg-0.16\deg$ (statistical only), meaning all sources were 
well localized. In the values reported in Table~1, we included a 10$\%$ 
allowance for systematic uncertainties as applied for the 2FGL catalog 
\cite{2fgl}.  This is a conservative estimate because the time intervals 
analyzed here are considerably shorter, and the statistical uncertainties 
correspondingly larger, than for the 2FGL catalog. The stellar counterparts for 
each nova were within the obtained LAT error circles, being within the $68\%$ 
confidence region for three cases, and within $95\%$ for \nsco\ which may suffer 
from larger systematic uncertainty due to large gradients in the bright diffuse 
\gray\ emission in the Galactic bulge.

\subsubsection*{\underline{Spectra}}

The LAT spectra of the novae were extracted in 12 logarithmic energy bins from 
100 MeV to 100 GeV. The fit results for single PL and exponentially cutoff 
power-law (EPL; the spectrum $N$($E$) $\propto E^{-s}e^{-E/E_{\rm c}}$) models 
are plotted over the LAT data in Fig.~S1 and summarized in Table~S1. With the 
extra degree of freedom in the EPL model versus the single PL, the fits were 
improved with significances of 3.4$\sigma$ for \ndel\ and $3.8\sigma$ for \nmon, 
but only marginally at 2.0$\sigma$ for \nsco. The new analysis of \ncyg\ yielded 
a 6.4$\sigma$ significance of the spectral curvature, confirming the original 
analysis in \cite{v407}. Interestingly, we also see marginal evidence in all 
four novae of a downturn of the LAT spectrum at the low energy end with no 
significant emission in the lowest energy (100$-$178 MeV) spectral point in each 
case. The $95\%$ confidence upper limits derived in this energy bin are at, or 
just below, the PL model for all four examples. Along with the statistical 
uncertainties, systematic uncertainties in the Galactic diffuse \gray\ emission 
limit the spectral shape measurements to varying degrees depending on the 
location of each nova studied. Because \nsco\ is seen against the Galactic bulge 
and has the brightest diffuse background, the spectral shape was more difficult 
to measure and the cutoff energy in the EPL fit was not well constrained. 
Regardless, the inferred cutoff energies are 1$-$4 GeV with indices = 1.2$-$1.8 
in the EPL model fits.

\subsubsection*{\underline{Search for Spectral Variability}}

In order to check whether the spectral characteristics changed during the \gray\ 
flares, each data set was split into the time intervals specified in Table~S1. 
The same method as the one presented previously for the full duration datasets 
was used to analyze both split sets of data. The resultant best-fit parameters 
are tabulated in Table~S1 and the spectral energy distributions are presented in 
Fig.~S2. For each nova, the first intervals (a) were six days long beginning at 
\t0, and resulted in comparable $TS$ values ($\sim$1$-$1.5$\times$ larger) than 
those in the second intervals (b) which range from 11 to 21 days long. Although 
the fluxes clearly decreased by factors of 1.6$-$2.6 between the two intervals, 
one cannot conclude that the spectral shape changed for any of the sources 
considering the measurement uncertainties. For \ncyg, these conclusions are 
again consistent with \cite{v407}.

\subsection*{S2.~High-energy Proton and Electron Spectra}

\medskip

\subsubsection*{\underline{Hadronic Model}}

Following \cite{v407}, the \gray\ spectra due to the decay of \p0\ produced in 
$pp$ collisions was calculated following the method presented in \cite{kam06s}, 
assuming a Solar metallicity \cite{mor09s}. We use high-energy proton spectra in 
the form $N_{\rm p} (p_{\rm p}) = N_{\rm p,0} \, (p_{\rm p} \, c)^{-s_{\rm p}} 
\, e^{-W_{\rm p}/E_{\rm cp}}$ (proton/GeV), where $p_{\rm p}$ is the proton 
momentum and $W_{\rm p}$ is the kinetic energy of protons [see, e.g., 
\cite{hou06}]. The parameters that were fitted to the data are $N_{\rm p,0}$, 
$s_{\rm p}$ and $E_{\rm cp}$ which are the normalization, the slope, and the 
cutoff energy of the high-energy proton spectrum, respectively. The fits were 
performed using the {\tt gtlike} likelihood fitting tool. The corresponding 
best-fit parameters and $TS$ values obtained for each nova are presented in 
Table~S1. The best-fit models are shown in Fig.~3 and the corresponding 
confidence regions of the slopes and cutoff energy fits are presented in 
Fig.~S3. These contour plots have similar shapes except the one for \ncyg, which 
shows a deep $\chi^{2}$ minimum at a slope of 1.4; one can also see a secondary 
$\chi^{2}$ minimum at about the value we published in \cite{v407}.

The total energy in high-energy protons can be calculated using estimates of the 
mean local target density over the time interval when the $\gamma$ rays were 
detected. In the hadronic model, we assumed that high-energy protons collide 
with the nuclei of the ejecta, except for V407 Cyg where the ejecta expand in 
the dense stellar wind of the companion star \cite{v407}. Consequently, in the 
latter case, high-energy protons collide with hydrogen nuclei that are in the 
ejecta and the wind. The high-energy protons are assumed to be uniformly 
distributed in the ejecta, which is transparent to $>$100 MeV $\gamma$ rays. For 
the three classical novae, the mean density of target hydrogen is assumed to 
correspond to the ejecta mass in the mean ejecta volume. Despite the likelihood 
that the ejecta are axisymmetric rather than spherical, for simplicity we 
calculated the mean ejecta volume over the \gray\ emission duration assuming an 
expanding shell with an outer radius $R_{\rm out}(t) = v_{\rm ej} \times t$ and 
a relative thickness $\Delta R / R_{\rm out} \approx 0.4$ [see, e.g., 
\cite{sho13,rib13}]. In these conditions, the total energy in high-energy 
protons ($\epsilon_{\rm p}$) resulting from the best fitting parameters ($N_{\rm 
p,0}$, $s_{\rm p}$, and $E_{\rm cp}$) is about $10^{42}-10^{43}$ ergs (see 
Table~S3) with proton lifetimes ranging from $\sim$ 100 to 300 days. The 
corresponding ratios $\eta_{\rm p}$ of the total energy in high-energy protons 
to the kinetic energy of the ejecta are presented with the ejecta parameters in 
Table~S3. The conversion efficiencies $\eta_{\rm p}$ in Table~S3 are similar to 
those estimated in supernova remnants [e.g., \cite{der13} and references 
therein]. The \gray\ emission produced by the secondary electrons and positrons 
from high-energy proton interactions with ejecta nuclei was not taken into 
account in this study since it requires more detailed modeling of the novae. 
This \gray\ emission component would contribute to the low energy range of the 
LAT spectra since the secondary lepton spectrum has a bump shape that peaks at 
$\sim$100 MeV. As the \gray\ emissivity is proportional to the number of target 
nuclei, the values of $\epsilon_{\rm p}$ and $\eta_{\rm p}$ presented in 
Table~S3 scale with the ejecta mass as $M_{\rm ej}^{-1}$ and $M_{\rm ej}^{-2}$, 
respectively.

\subsubsection*{\underline{Leptonic Model}}

In the leptonic model, the inverse Compton (IC) and bremsstrahlung spectra are 
calculated using the method presented in \cite{blu70s}, for a high-energy 
electron spectrum that is an exponentially cutoff power-law: $N_{\rm e}(W_{\rm 
e}) = N_{\rm e,0} W_{\rm e}^{-s_{\rm e}} e^{-W_{\rm e} / E_{\rm ce}}$ 
(electron/GeV), where $W_{\rm e}$ is the kinetic energy of electrons. As in the 
hadronic model, the normalization ($N_{\rm e,0}$), the slope ($s_{\rm e}$) and 
the cutoff energy ($E_{\rm ce}$) were fitted to the LAT data with the {\tt 
gtlike} likelihood fitting tool. The results of the fit for each nova are 
presented in Table~S1 and the corresponding confidence regions of the slope and 
cutoff energy fit in Fig.~S4. The best-fit models are presented in Fig.~3.

The target photons used to calculate the \gray\ spectrum produced by the IC 
process are emitted by the nova photosphere. This differs from the first study 
of the \ncyg\ \gray\ emission \cite{v407} for which the target photons were 
assumed to be emitted by the red giant (RG) companion star. The target photon 
spectrum was modeled by a black body with a temperature of 15 000 K and a radius 
of 3$\times$10$^{12}$ cm, taken as typical values characterizing nova 
photospheres. The photon distribution before scattering is assumed to be 
isotropic in the high energy electron frame. The high-energy electrons were 
assumed to be at the ejecta front layer, at a distance from the white dwarf that 
corresponds to the average radius of the ejecta during the \gray\ emission 
duration, taking into account the ejecta velocity ($v_{\rm ej}$) presented in 
Table~S3. In order to estimate the effects of high-energy electron interactions 
with the atoms in the ejecta, we calculated the bremsstrahlung spectrum assuming 
that high-energy electrons interact with a canonical ejecta mass of $10^{-5} 
M_{\odot}$ (Solar metallicity), except for \ncyg\ where the additional target 
atoms in the dense stellar wind of the companion red giant star were taken into 
account \cite{v407}. The separate contributions of the IC and bremsstrahlung 
spectral components in the best-fit leptonic models are presented in Fig.~S5. 
The model parameters fitted with the LAT data led to the total high-energy 
electron energies and conversion efficiencies presented in Table~S4. The 
resulting electron lifetimes range from $\sim$4 to 10 days. The conversion 
efficiencies were calculated assuming kinetic energies derived from ejecta 
parameters presented in Table~S3. It should be noted that the values of the 
total energy of high-energy electrons obtained from the fits scale with the 
inverse of the nova luminosity since more target photons implies fewer 
high-energy electrons are needed to produce the same \gray\ luminosity.

\subsection*{S3.~Notes on Available Optical Observations of the Novae}

\medskip

\subsubsection*{\underline{V407 Cyg 2010}}

The symbiotic binary system \ncyg\ was studied in detail by \cite{mun90}, who 
found that the secondary is a Mira variable with spectral type M6 III and 
distance of 2.7 kpc as assumed here and in our previous LAT \gray\ detection 
paper on the 2010 nova \cite{v407}.

\subsubsection*{\underline{V959 Mon 2012}}

The historical optical and near-infrared counterpart of Nova Monocerotis 2012 
(\nmon) was found in archival databases by \cite{gre12s}.  These consisted of 
optical observations ($r \simeq 17.8 - 18.1$ and $i = 17.1 - 17.4$ mag) from the 
INT Photometric H-Alpha Survey that used the 2.5m Isaac Newton Telescope and 
near-infrared photometry ($J = 16.26$, $H = 15.71$, and $K = 15.42$ mag) from 
the UKIDSS survey that used the 3.8m UK Infrared Telescope. Considering a 
plausible range of values of the extinction $E(B-V)=0.3-0.8$ [see also 
\cite{sho13}], \cite{gre12s} concluded that the companion is likely a 
main-sequence star.

\subsubsection*{\underline{V1324 Sco 2012}}

The available multi-wavelength observations of Nova Scorpii 2012 (\nsco) are 
relatively sparse compared to \nmon\ and \ndel. In the report of its optical 
discovery, \cite{wag12} found a variable pre-outburst counterpart with $I = 
19.0-19.5$ mag. The AAVSO light curve shows a $V$-band maximum around 10th mag.  
The $B$ magnitude at maximum was about 11.5 which, assuming an intrinsic 
$(B-V)_{\rm 0} \sim 0$ [e.g., \cite{van87s}], implies a large $E(B-V)$ value of 
1.5 mag.  With a time of decline by two magnitudes $t_{\rm 2}$ of about 25 days 
estimated from the AAVSO light curve, we obtain an absolute $V$-band magnitude 
of $-7.6$ from the maximum magnitude rate of decline relation of \cite{del95s} 
and a distance of 4 kpc. We emphasize the uncertainty in this method 
\cite{kas11} and adopted it in the absence of any more direct method for 
obtaining the intrinsic properties. Note, however, that the substantial drop in 
brightness might be from a dust-forming event that compromises the $t_{\rm 2}$ 
value; without dust formation the $t_{\rm 2}$ time would be much longer and the 
inferred source closer. The same applies if the extinction has been 
overestimated.  For example if $E(B-V) \sim 1$ mag, which might be supported by 
the low neutral hydrogen column density toward this source, the distance is 8 
kpc. Given the uncertainties outlined, we adopted the distance of 4.5 kpc from 
\cite{hil12} and the absolute magnitude of the quiescent counterpart in the 
$I$-band is $\sim 2$ mag.

\subsubsection*{\underline{V339 Del 2013}}

Nova Delphini 2013 (\ndel) was discovered as PNV J20233073+2046041 by Koichi 
Itagaki (Teppo-cho, Yamagata, Japan) at 6.8 magnitude on 2013 August 14.584 UT 
with nothing visible in an image taken with a limiting magnitude of 13.0 on 
August 13.565 \cite{ita13}. Optical spectroscopy identified it as a classical 
nova \cite{sho13del1,dar13s,tom13s}. Pre-discovery images indicated a rapid 
brightening from 10.1 to 8.5 magnitude from 7hr UT to 8hr UT on the discovery 
date \cite{wre13s}. The historical $V = 17.1$ mag of the optical counterpart 
comes from APASS observations \cite{mun13s}; see also \cite{dea14s}. With the 
small extinction $E(B-V) \simeq 0.2$ \cite{mun13s2,sho13s2,tom13s}, the absolute 
magnitude is 3.5 mag making this unlikely a red giant. We obtained $V$-band 
magnitudes from the American Association of Variable Star Observers 
(AAVSO)\footnote{\url{http://www.aavso.org/lcg}} \cite{aavso}, averaged them in 
0.25 day bins, and these are plotted along with the \gray\ light curves 
(Fig.~S6). The naked-eye peak visual 4.3 magnitude, also from the AAVSO, was 
observed on 2013 August 16.5, two days after discovery.


\newpage
\begin{figure}[htbp]
  \begin{center}
    \includegraphics[width=7.75cm]{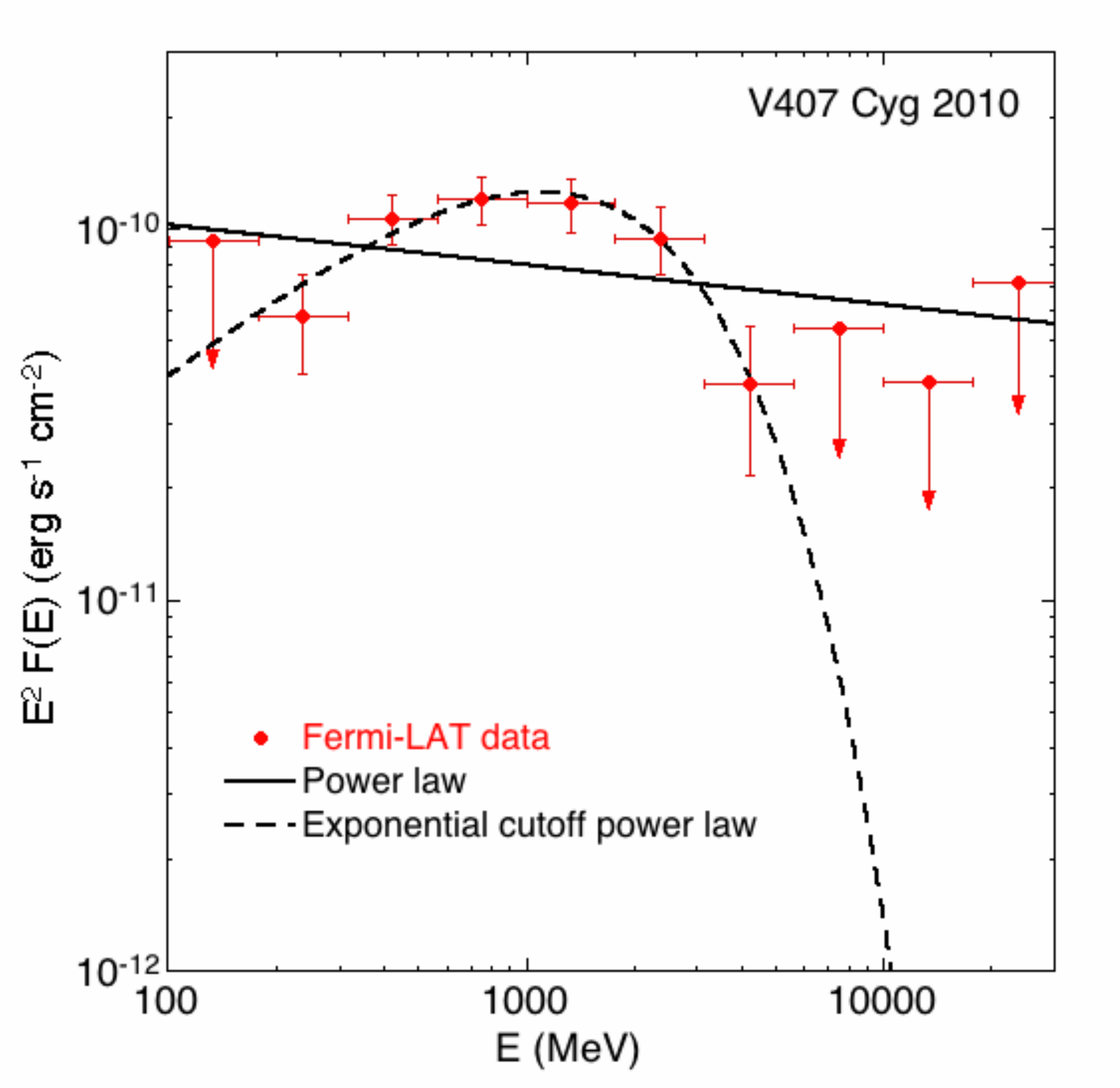}
    \includegraphics[width=7.75cm]{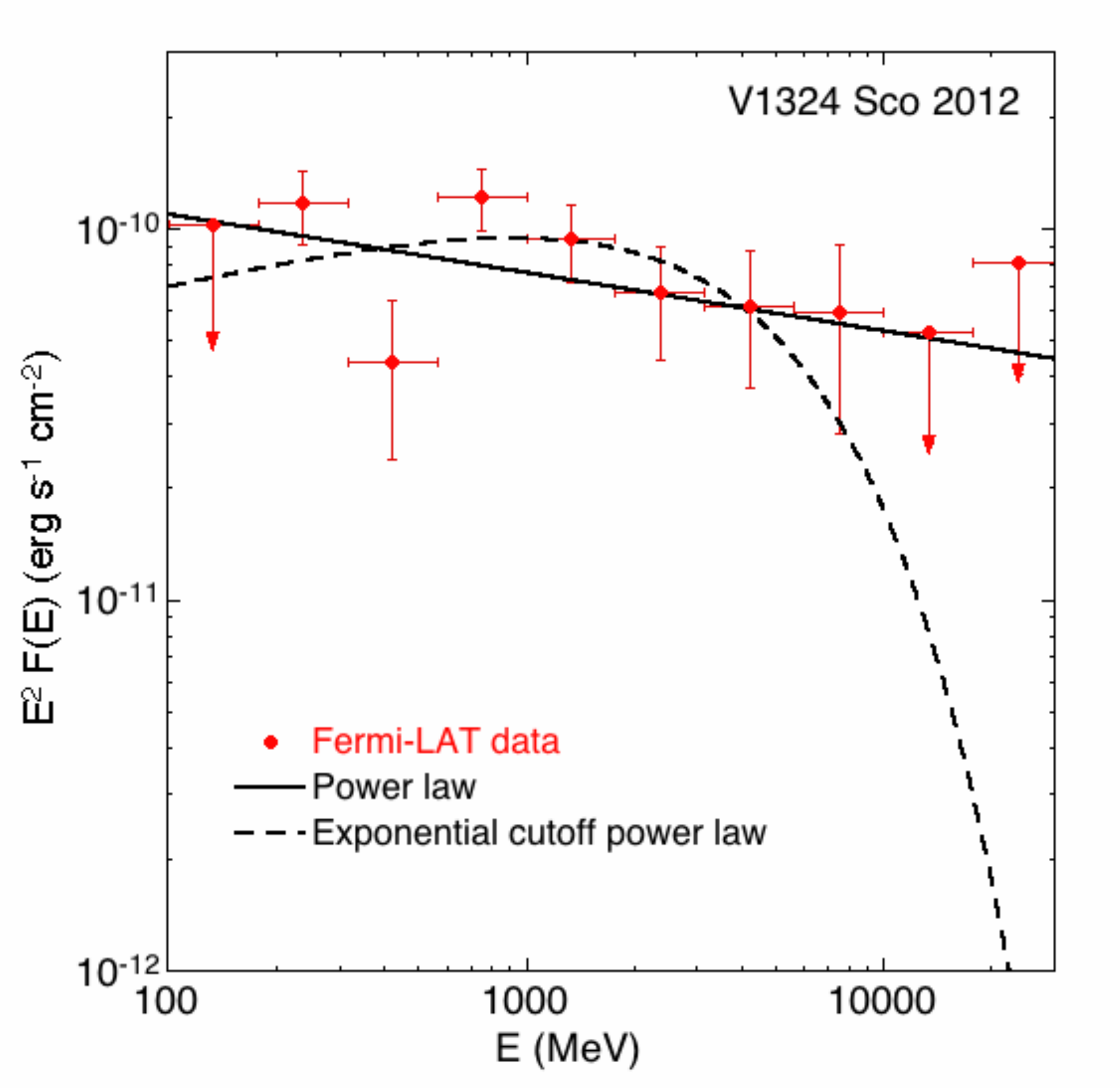}
    \includegraphics[width=7.75cm]{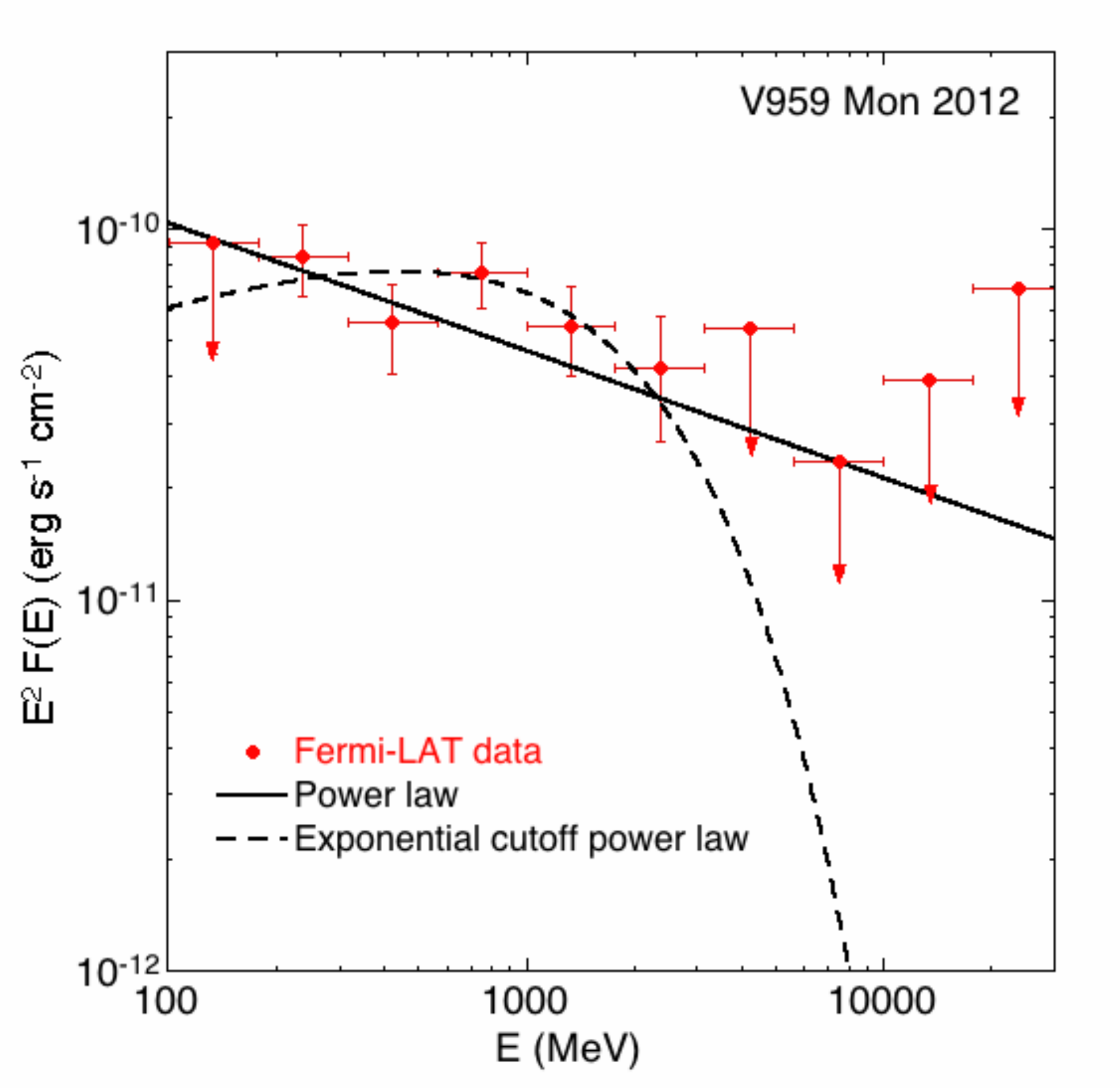}
    \includegraphics[width=7.75cm]{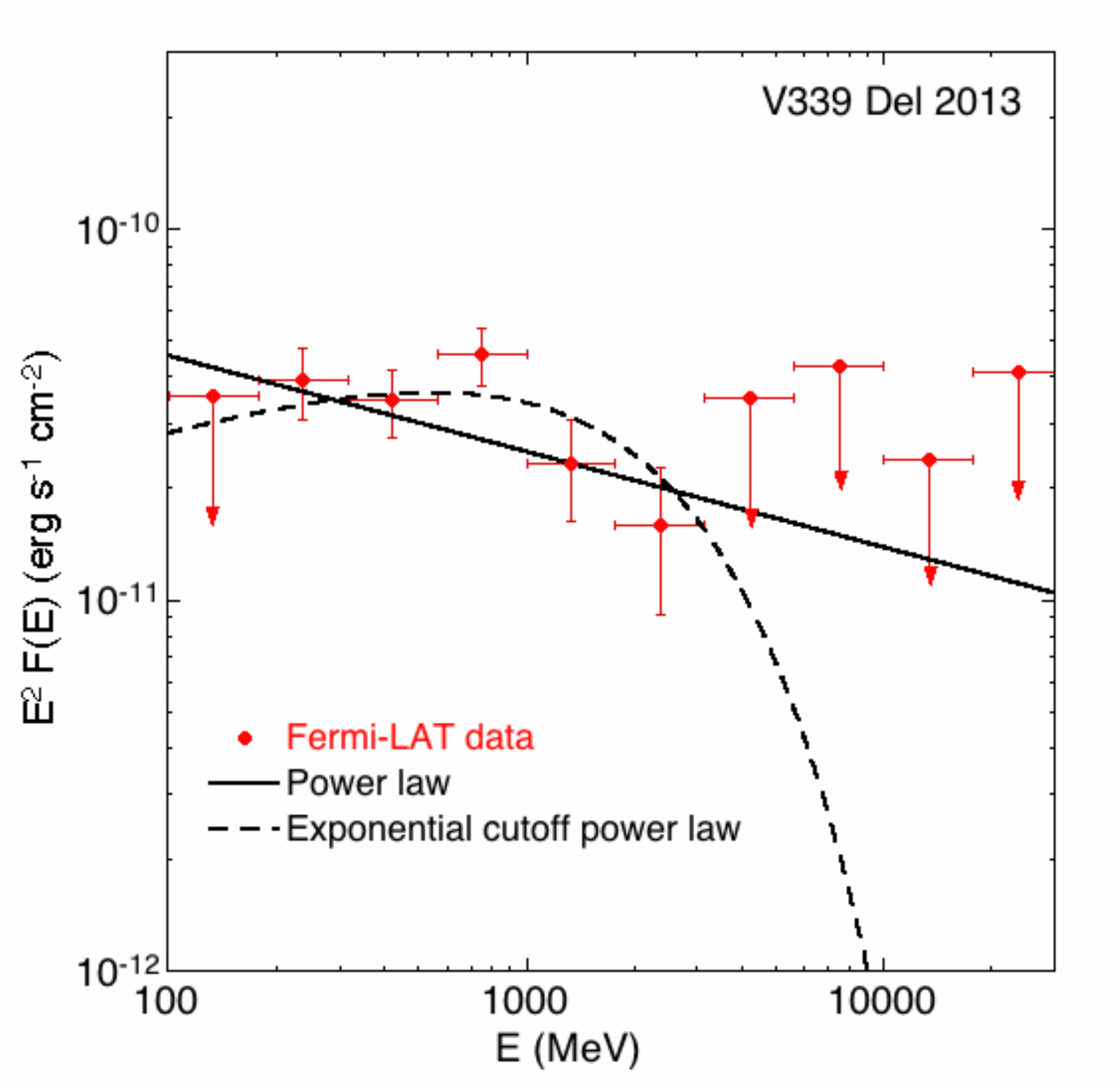}
  \end{center}
{\bf Fig.~S1.}
\Fermi-LAT $>$100 MeV average \gray\ spectra of the four novae over the full
17$-$27 day durations. Vertical bars indicate 1$\sigma$ uncertainties for data
points with $TS > 4$; otherwise, arrows indicate 2$\sigma$ limits. The best-fit
PL and EPL models (Table~S1) are overlaid on the LAT data.
\end{figure}

\newpage
\begin{figure}[htbp]
  \begin{center}
    \includegraphics[width=7.75cm]{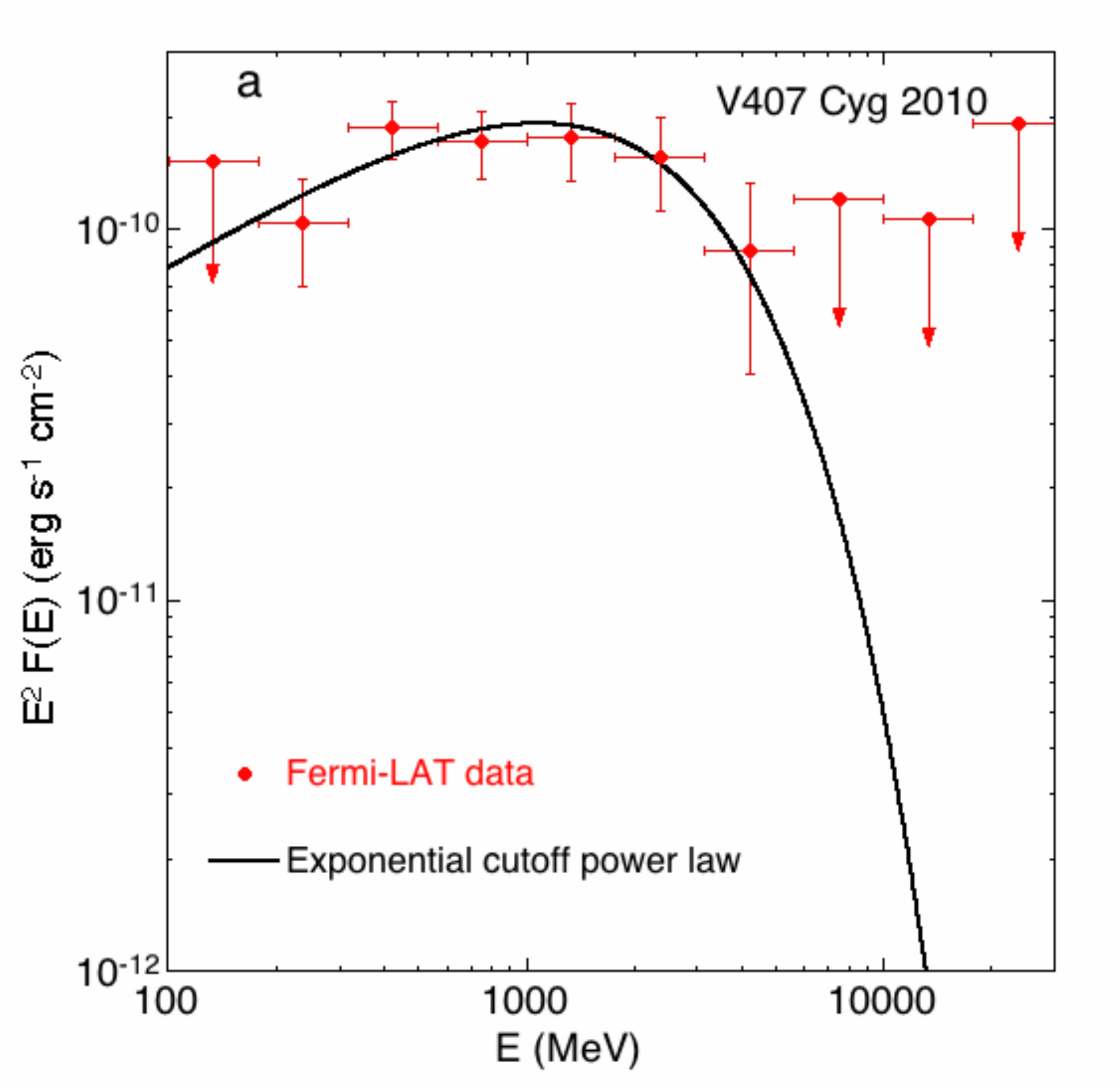}\includegraphics[width=7.75cm]{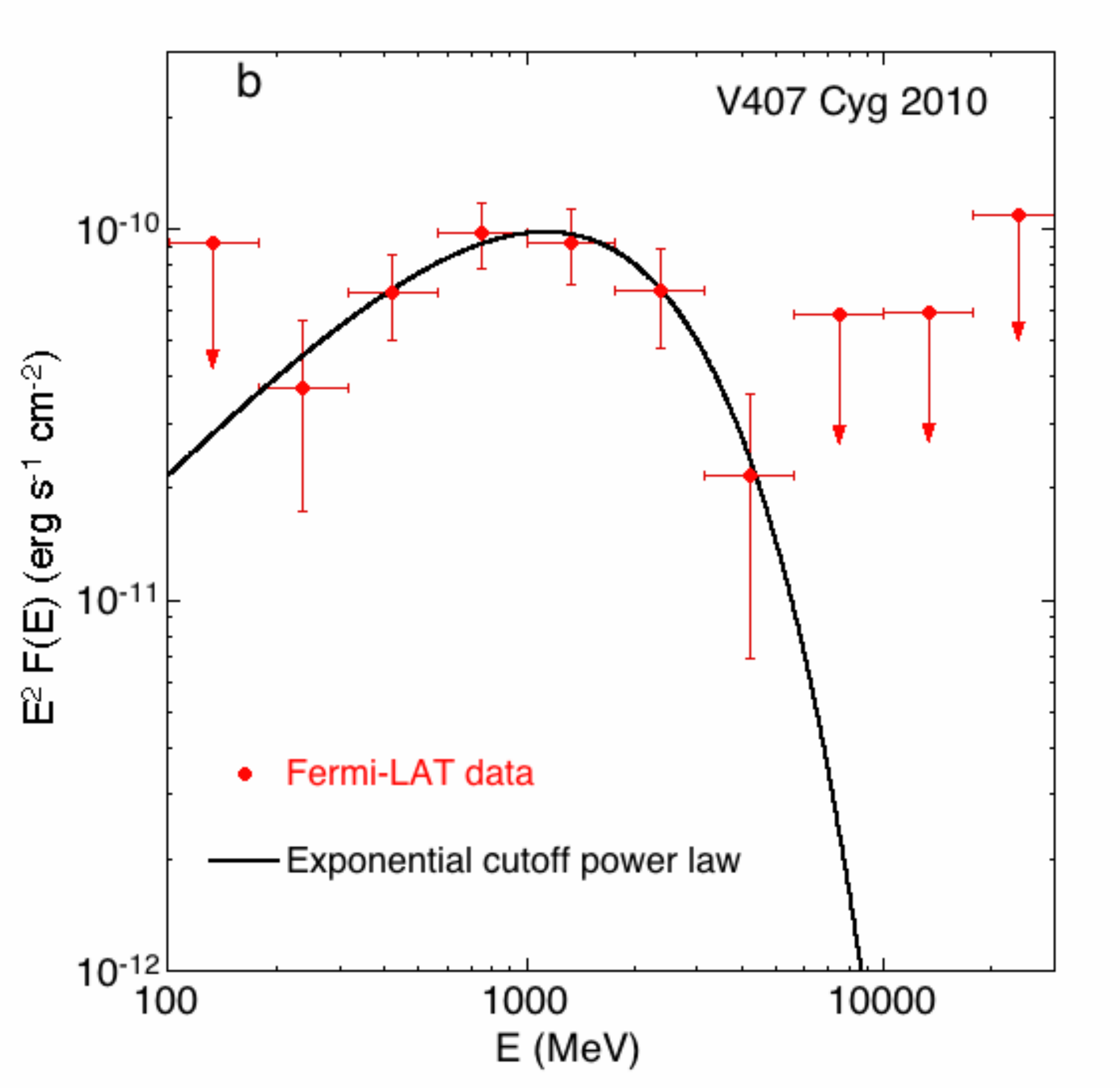}
    \includegraphics[width=7.75cm]{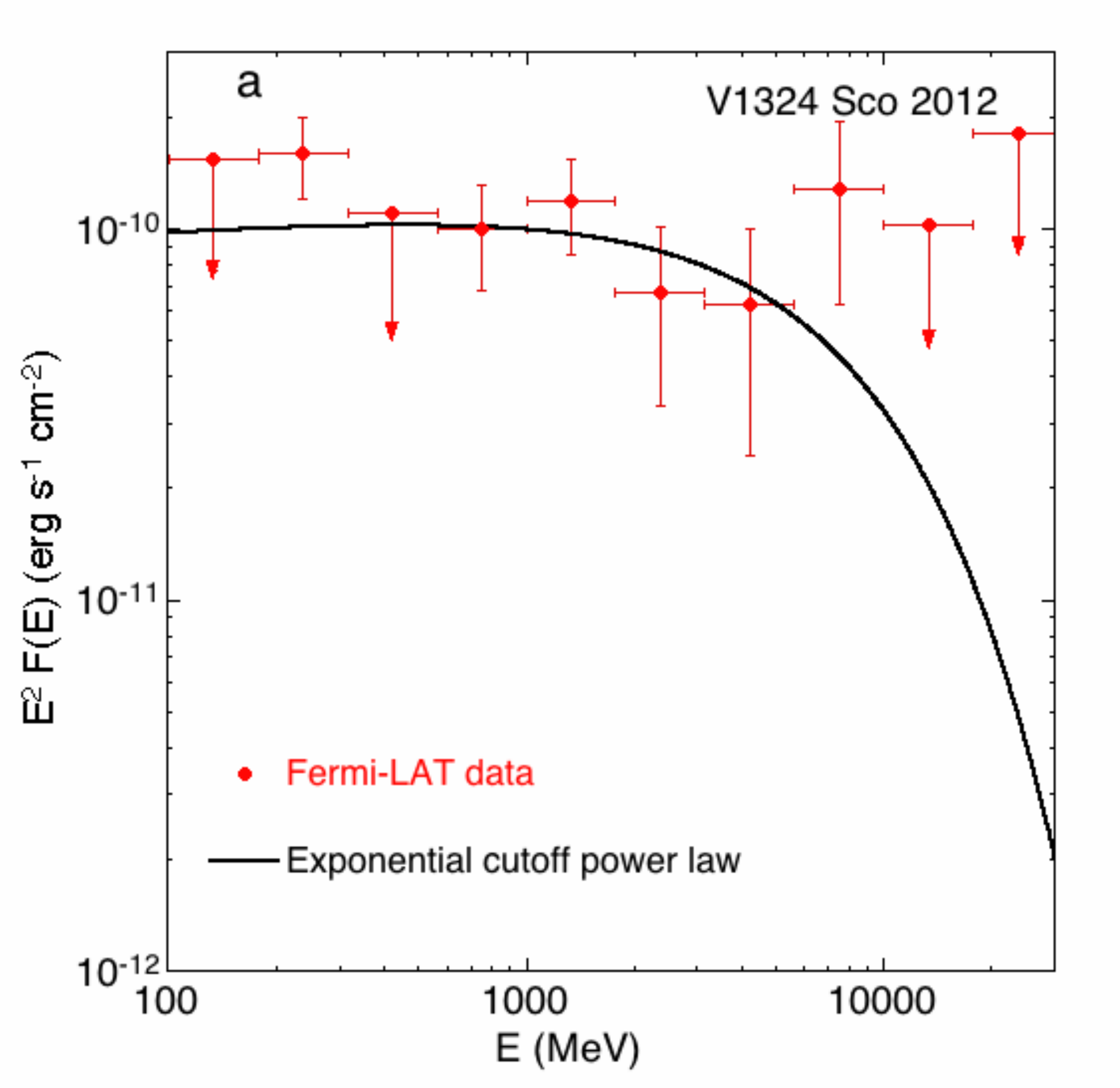}\includegraphics[width=7.75cm]{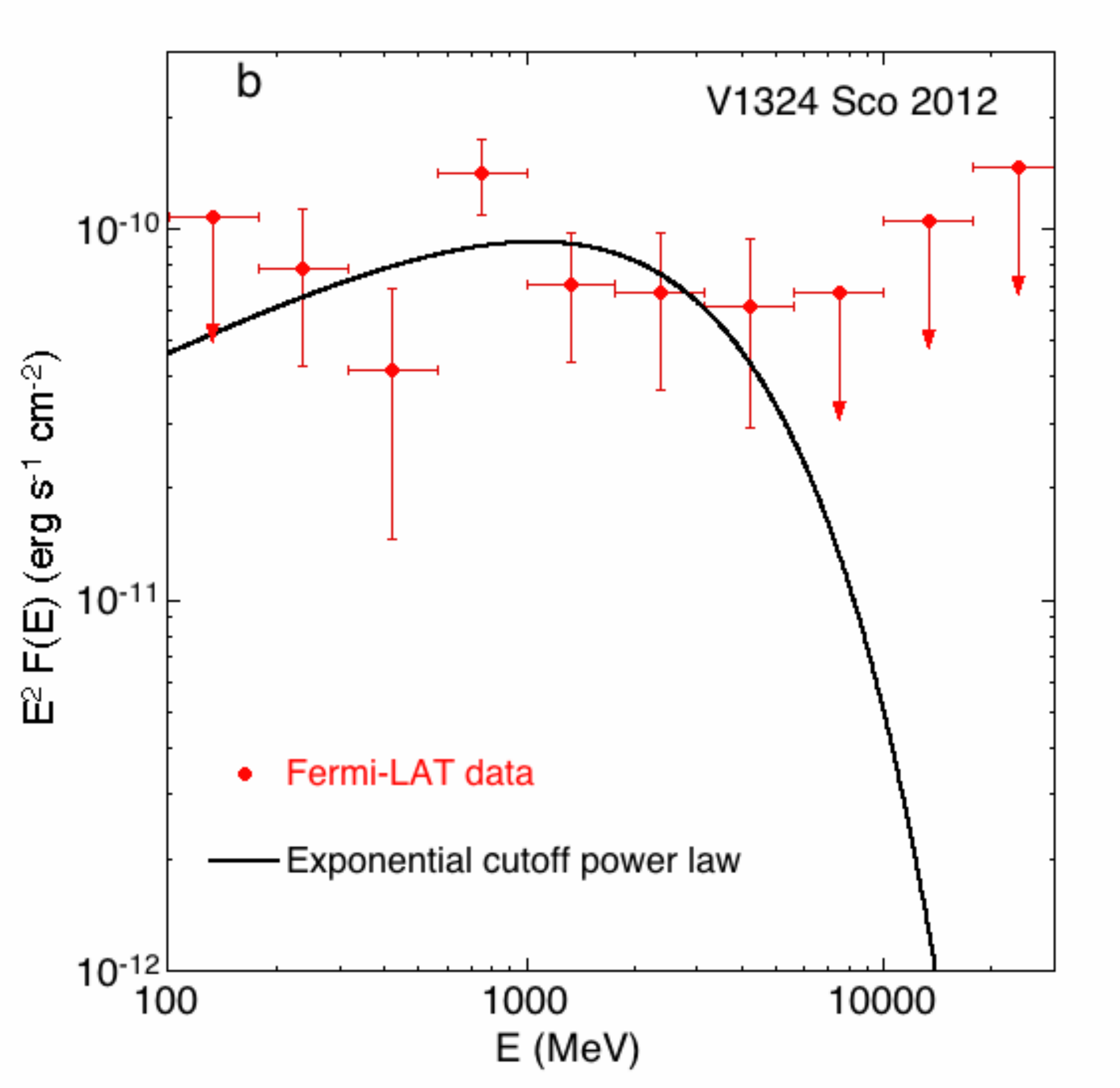}
  \end{center}
{\bf Fig.~S2.}
 \Fermi-LAT $>$100 MeV average \gray\ spectra of
the four novae as in Fig.~S1 but split into two intervals denoted as panels (a)
and (b). Vertical bars indicate 1$\sigma$ uncertainties for data points with $TS
> 4$; otherwise, arrows indicate 2$\sigma$ limits. The best-fit EPL models are
overlaid onto the LAT data. See Table~S1 for details.
\end{figure}

\newpage
\begin{figure}[htbp]
  \begin{center}
    \includegraphics[width=7.75cm]{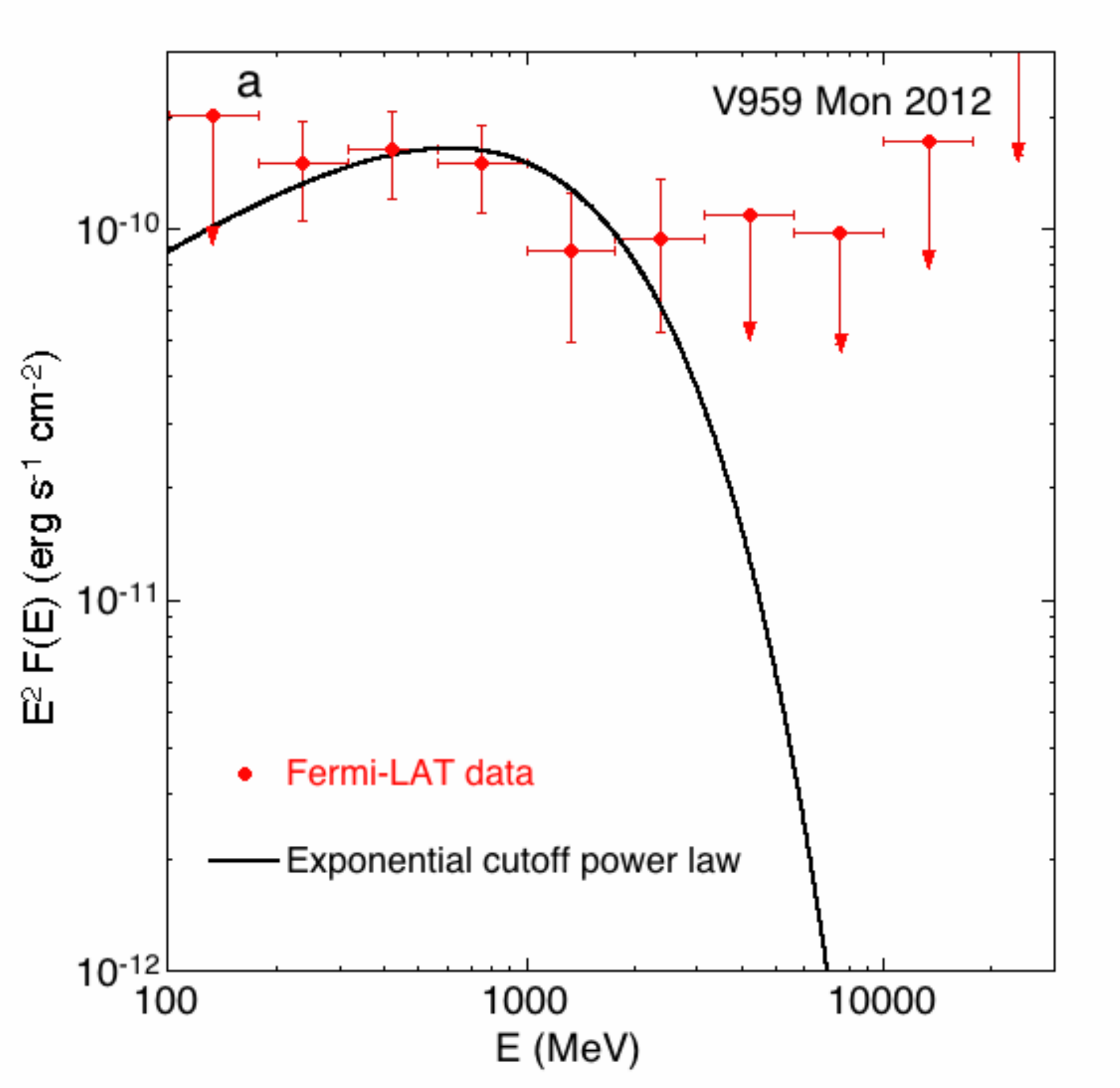}\includegraphics[width=7.75cm]{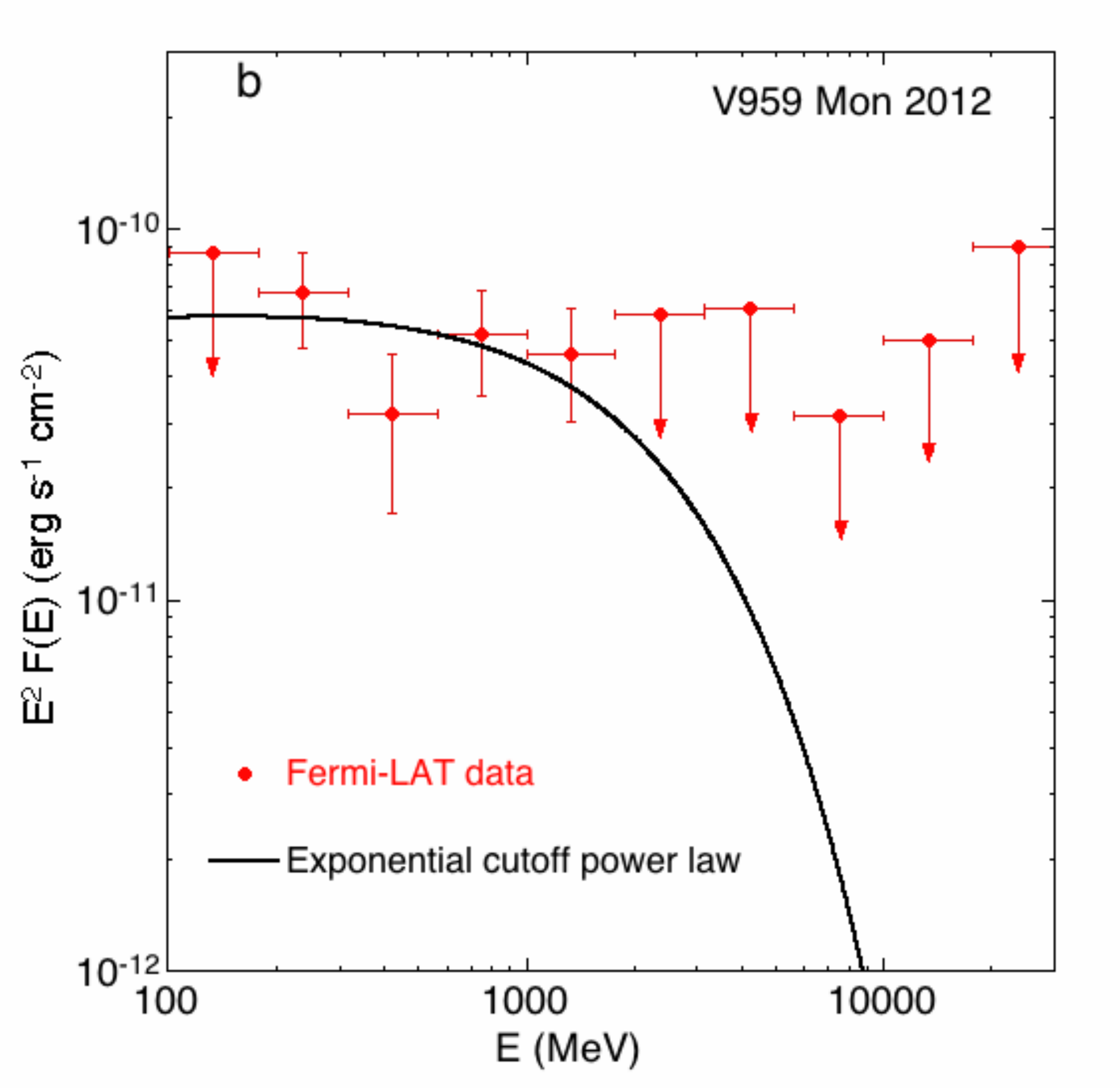}
    \includegraphics[width=7.75cm]{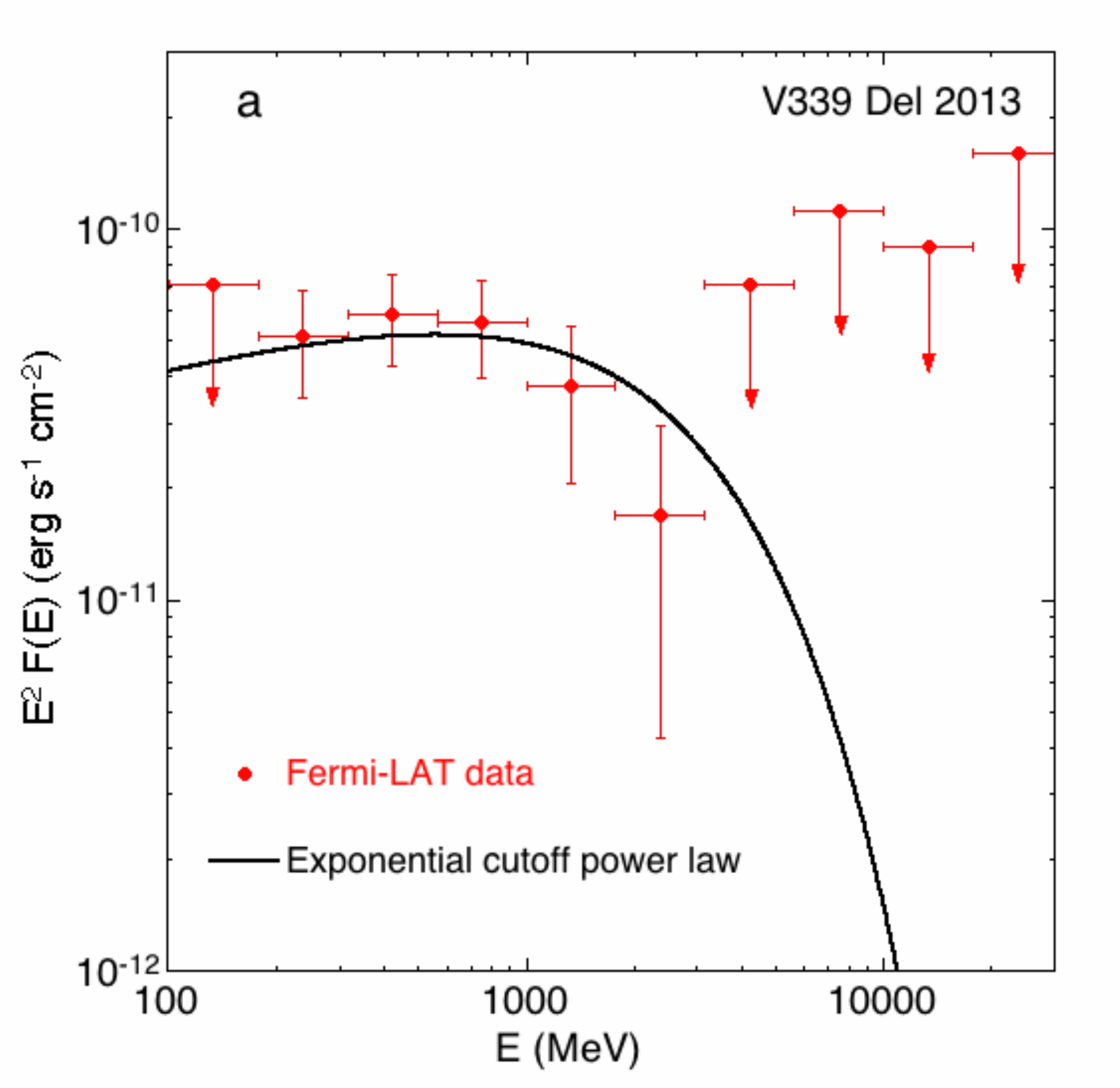}\includegraphics[width=7.75cm]{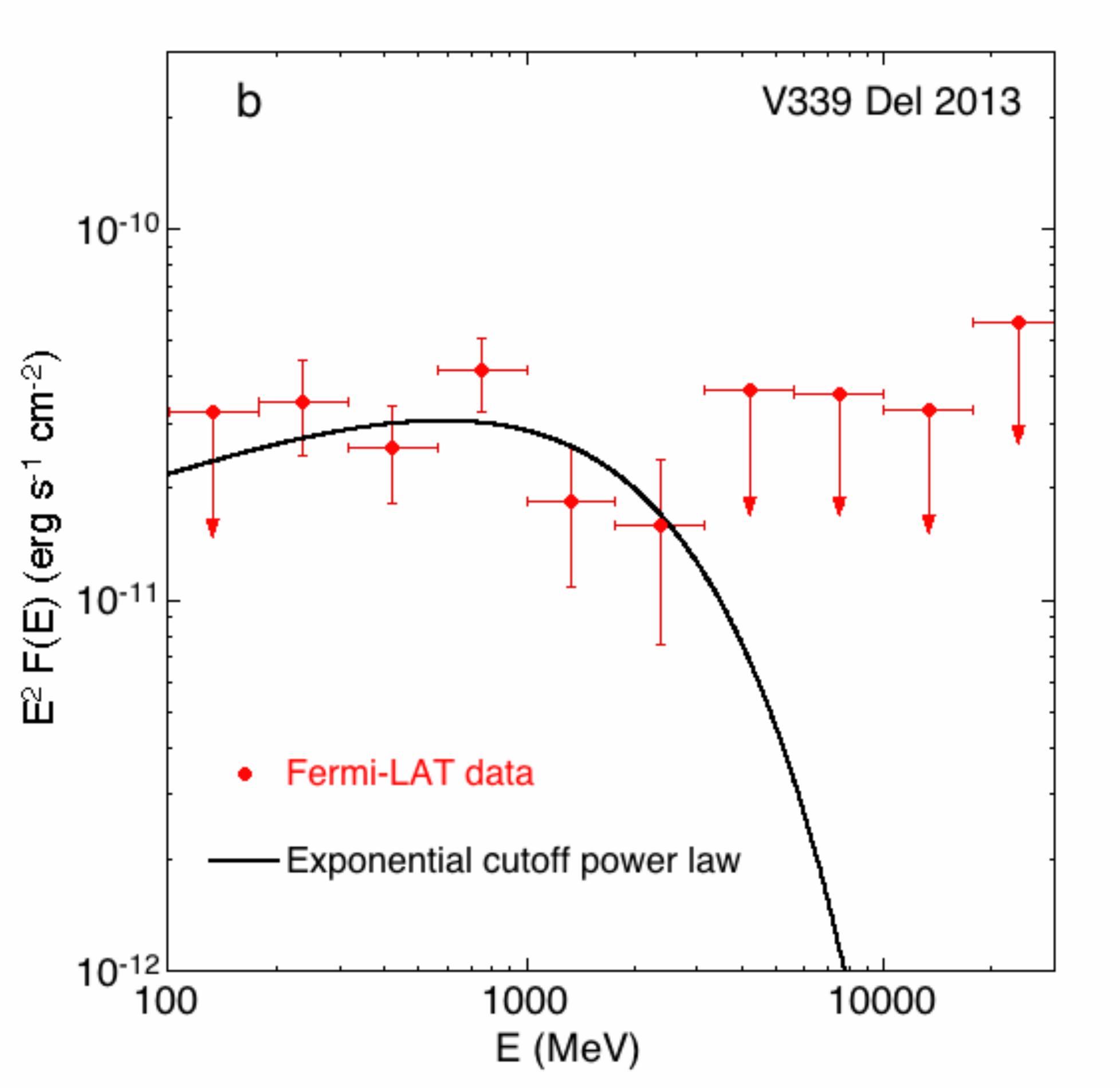}
  \end{center}
{\bf Fig.~S2.}
{\it continued.}
\end{figure}

\newpage
\begin{sidewaysfigure}[htbp]
  \begin{center}
    \includegraphics[width=6.5cm,angle=90]{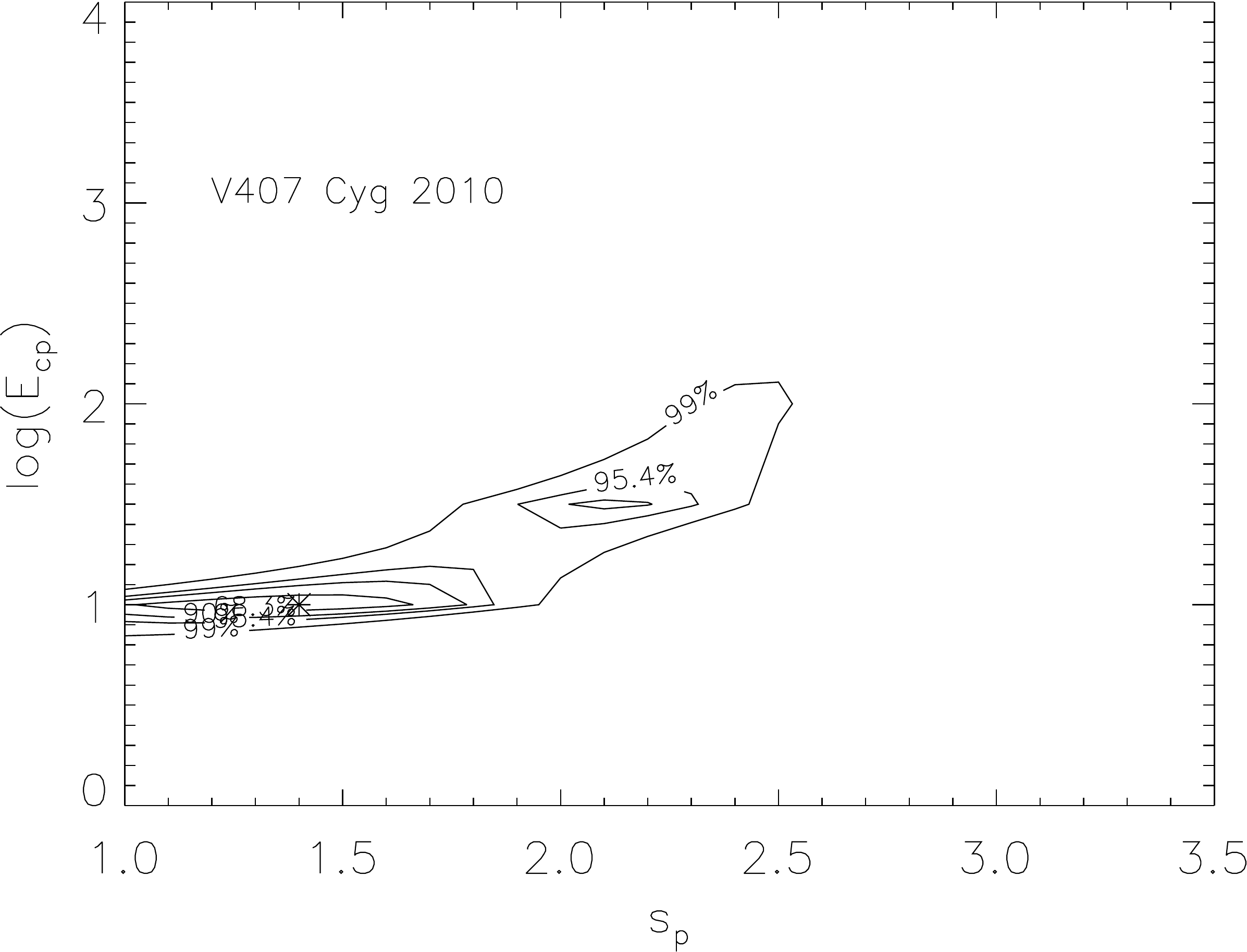}
    \includegraphics[width=6.5cm,angle=90]{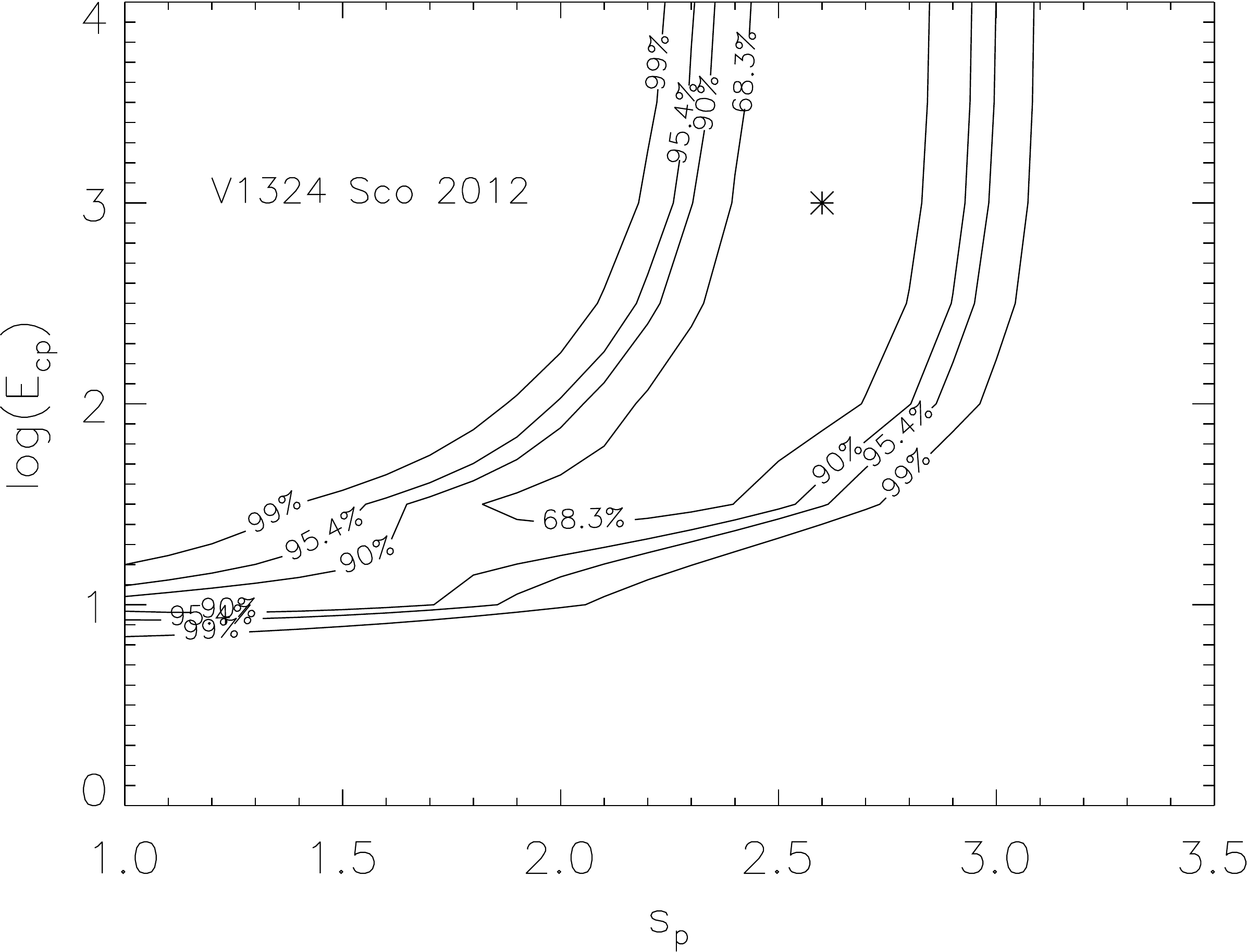}
    \includegraphics[width=6.5cm,angle=90]{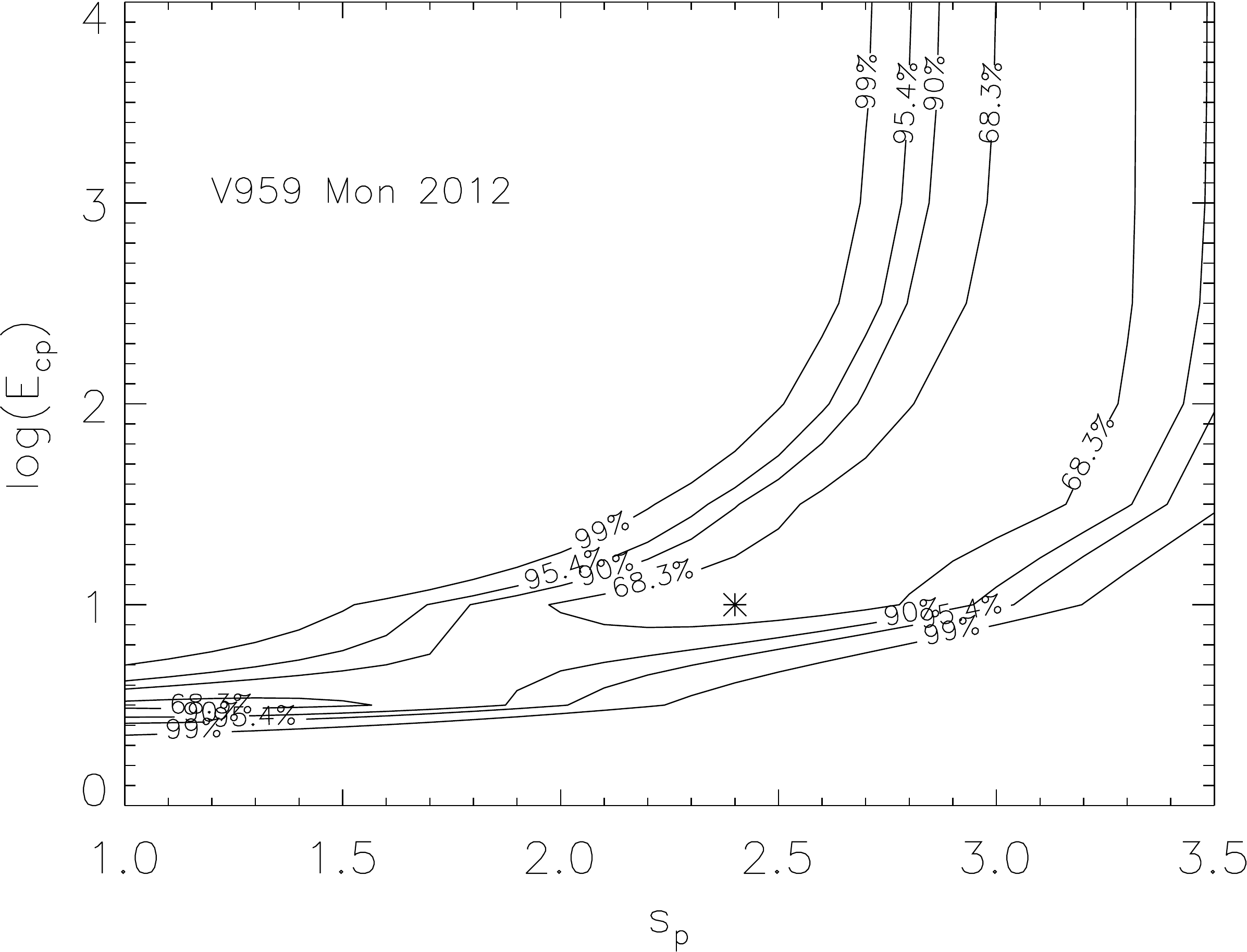}
    \includegraphics[width=6.5cm,angle=90]{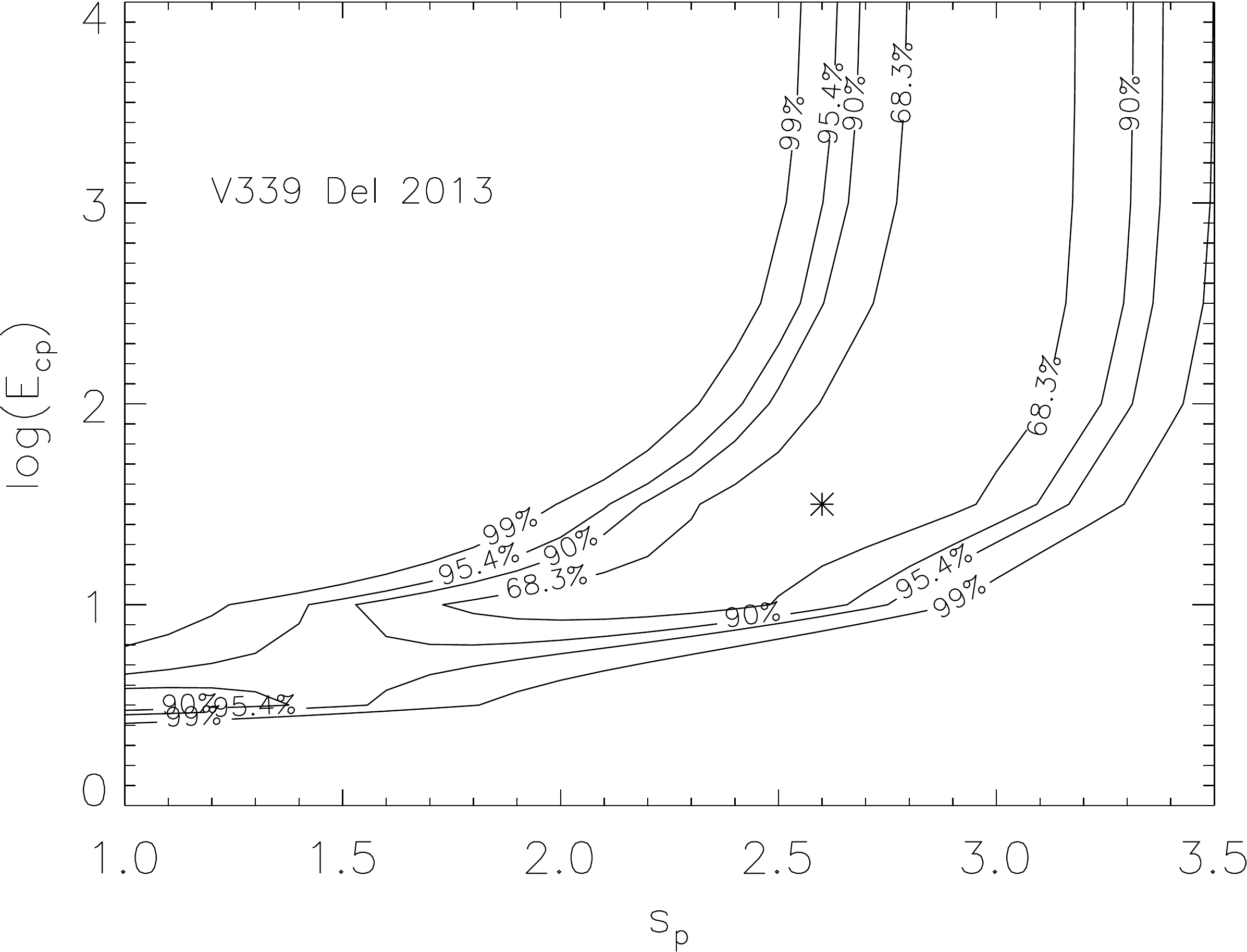}
  \end{center}
{\bf Fig.~S3.}
Contour plots of the hadronic model fits of the high-energy proton spectral
parameters for the four novae (energies $E$ in units of GeV). The star symbol in
each panel indicates the best-fit value (see Table~S1).
\end{sidewaysfigure}

\newpage
\begin{sidewaysfigure}[htbp]
  \begin{center}
    \includegraphics[width=6.5cm,angle=90]{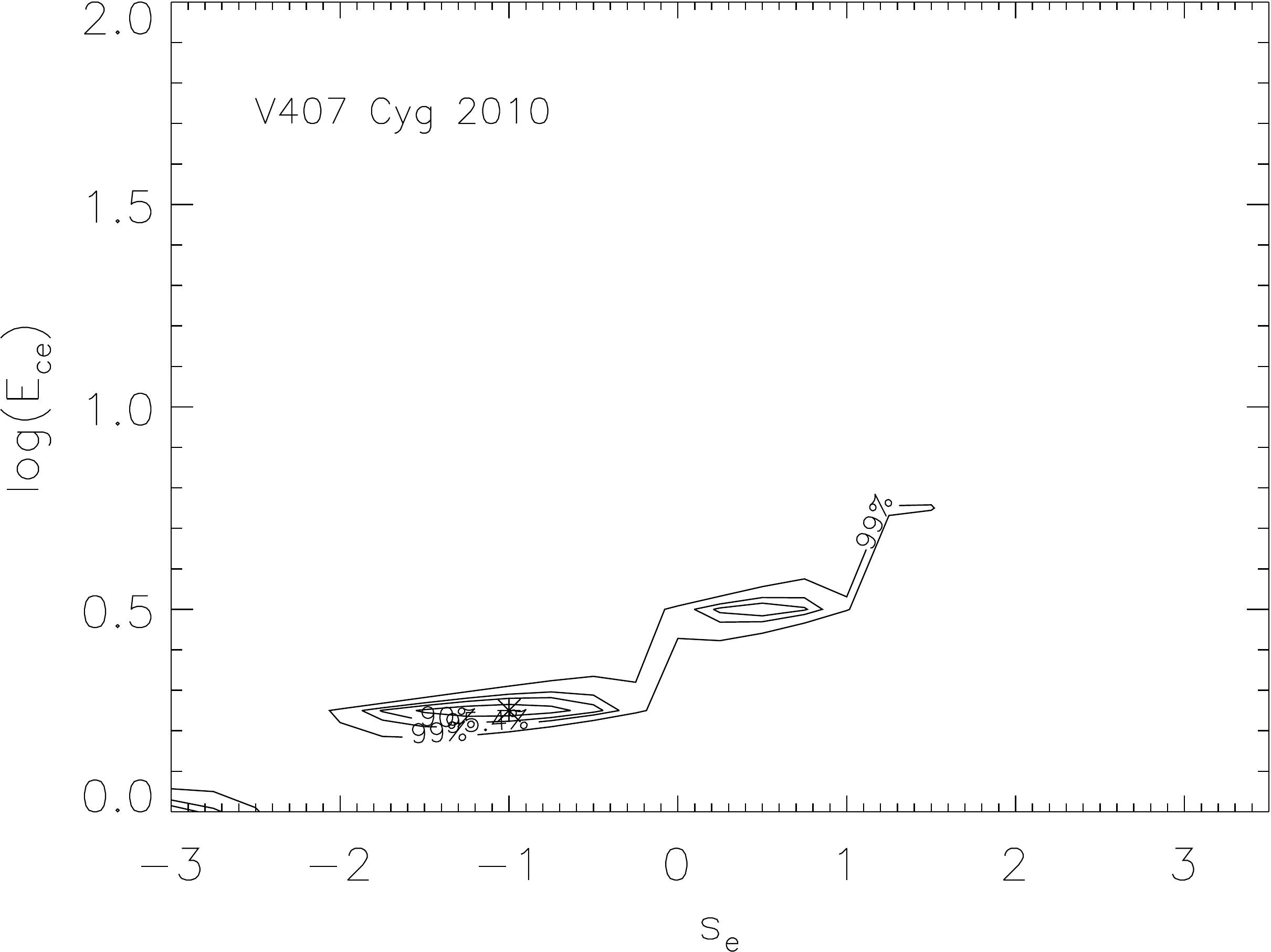}
    \includegraphics[width=6.5cm,angle=90]{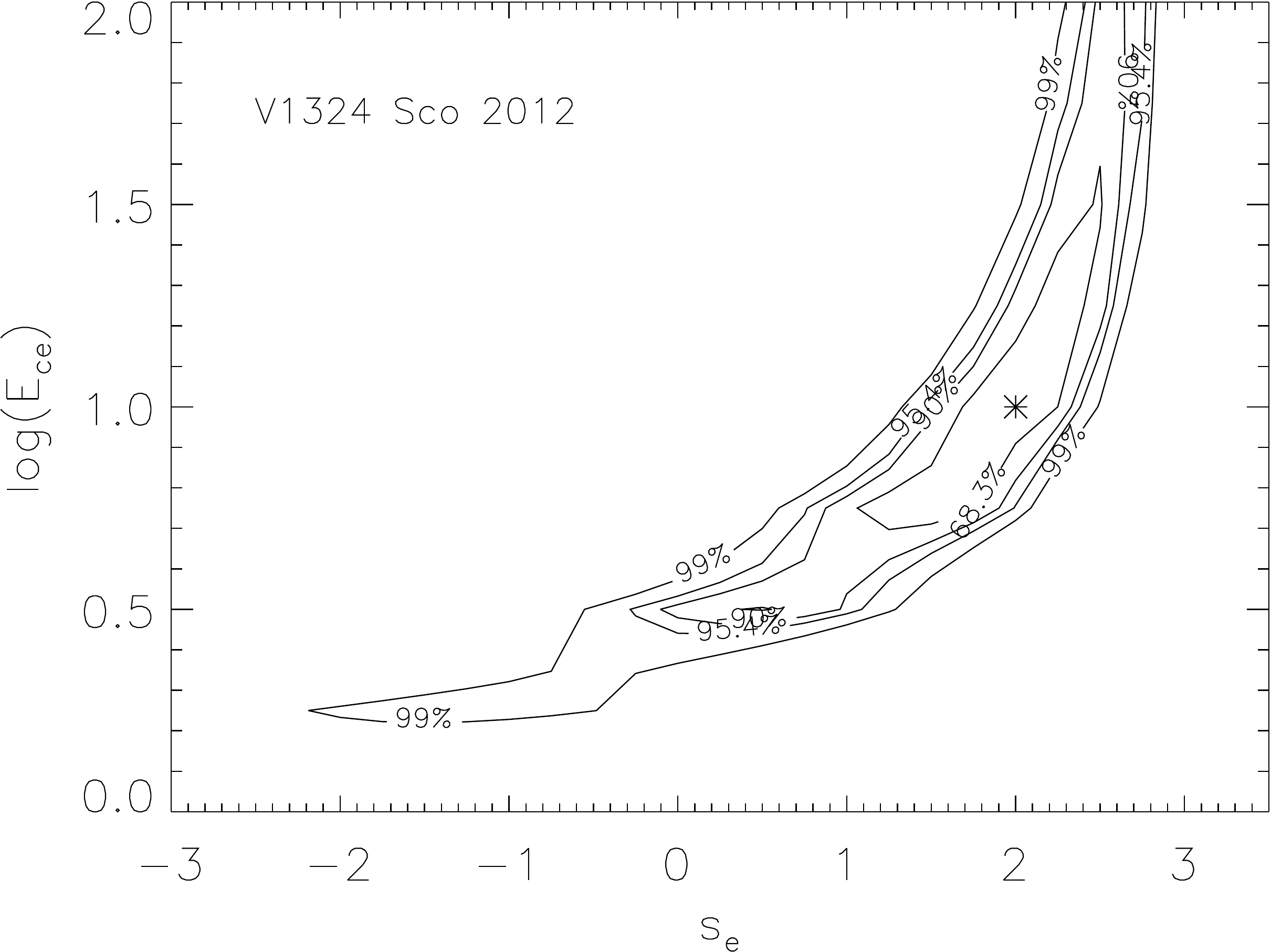}
    \includegraphics[width=6.5cm,angle=90]{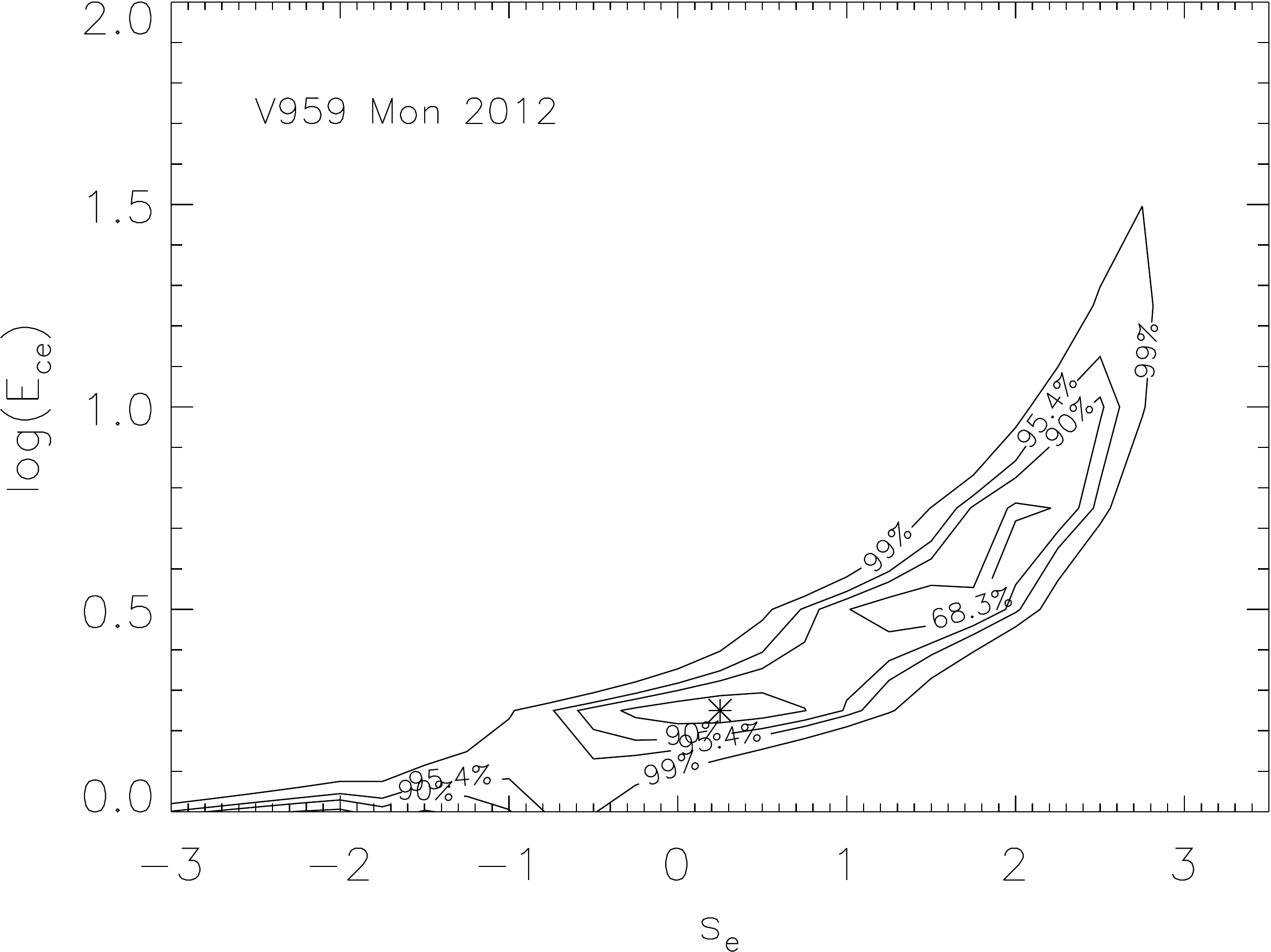}
    \includegraphics[width=6.5cm,angle=90]{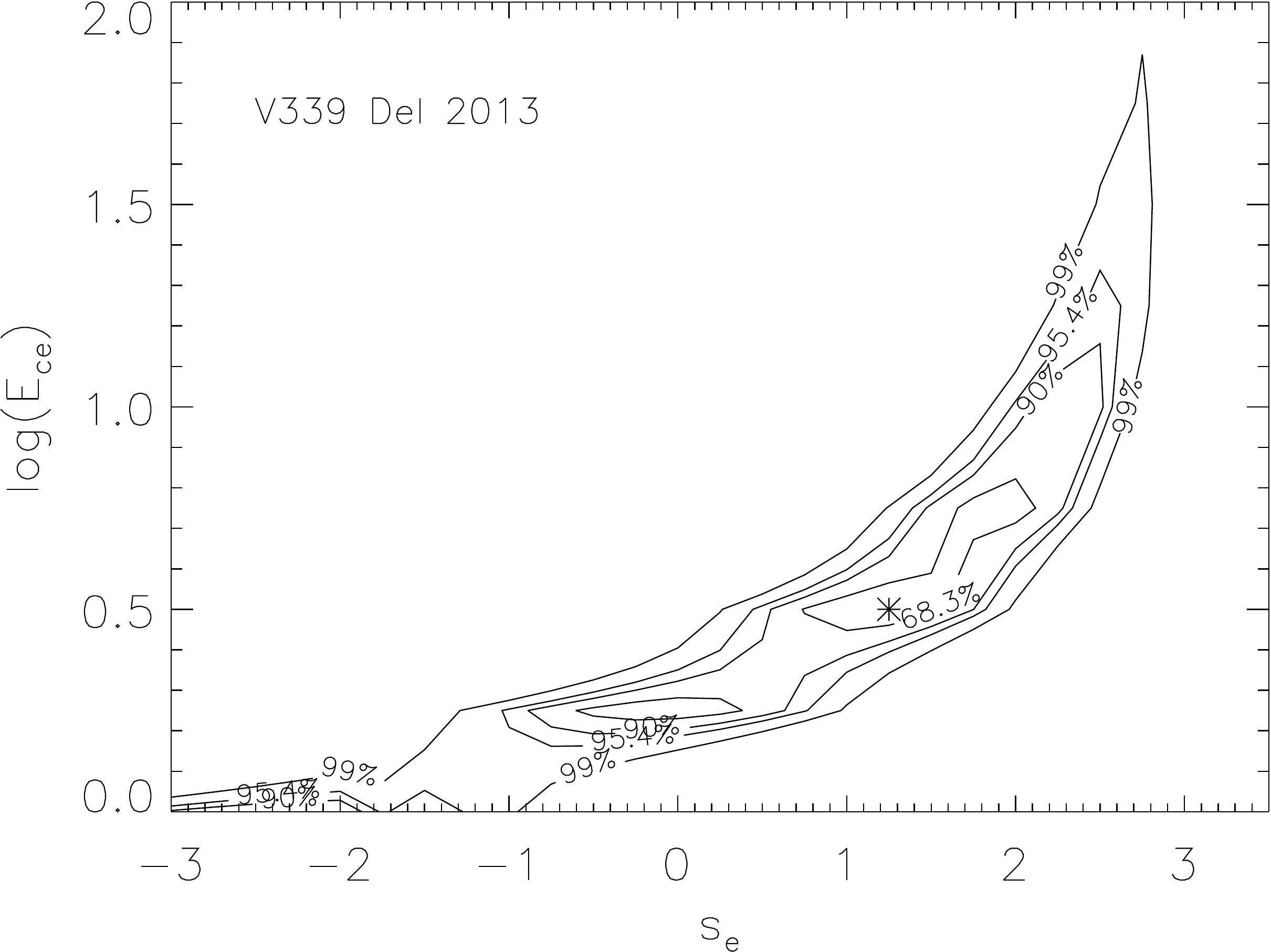}
  \end{center}
{\bf Fig.~S4.}
As in Fig.~S3, but for the leptonic model fits of the high-energy electron
spectral parameters.
\end{sidewaysfigure}

\newpage
\begin{figure}[htbp]
  \begin{center}
    \includegraphics[width=7.75cm]{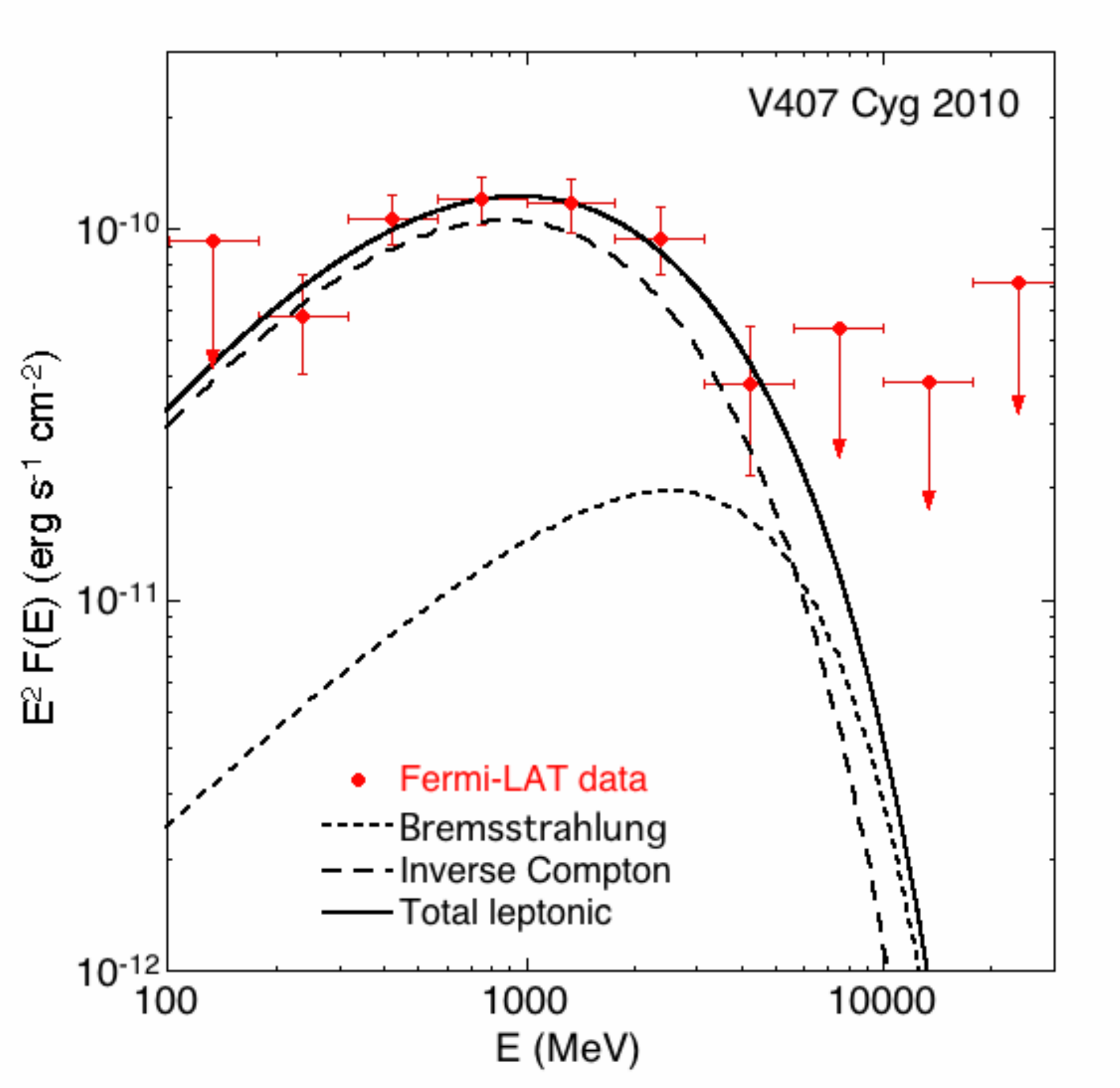}
    \includegraphics[width=7.75cm]{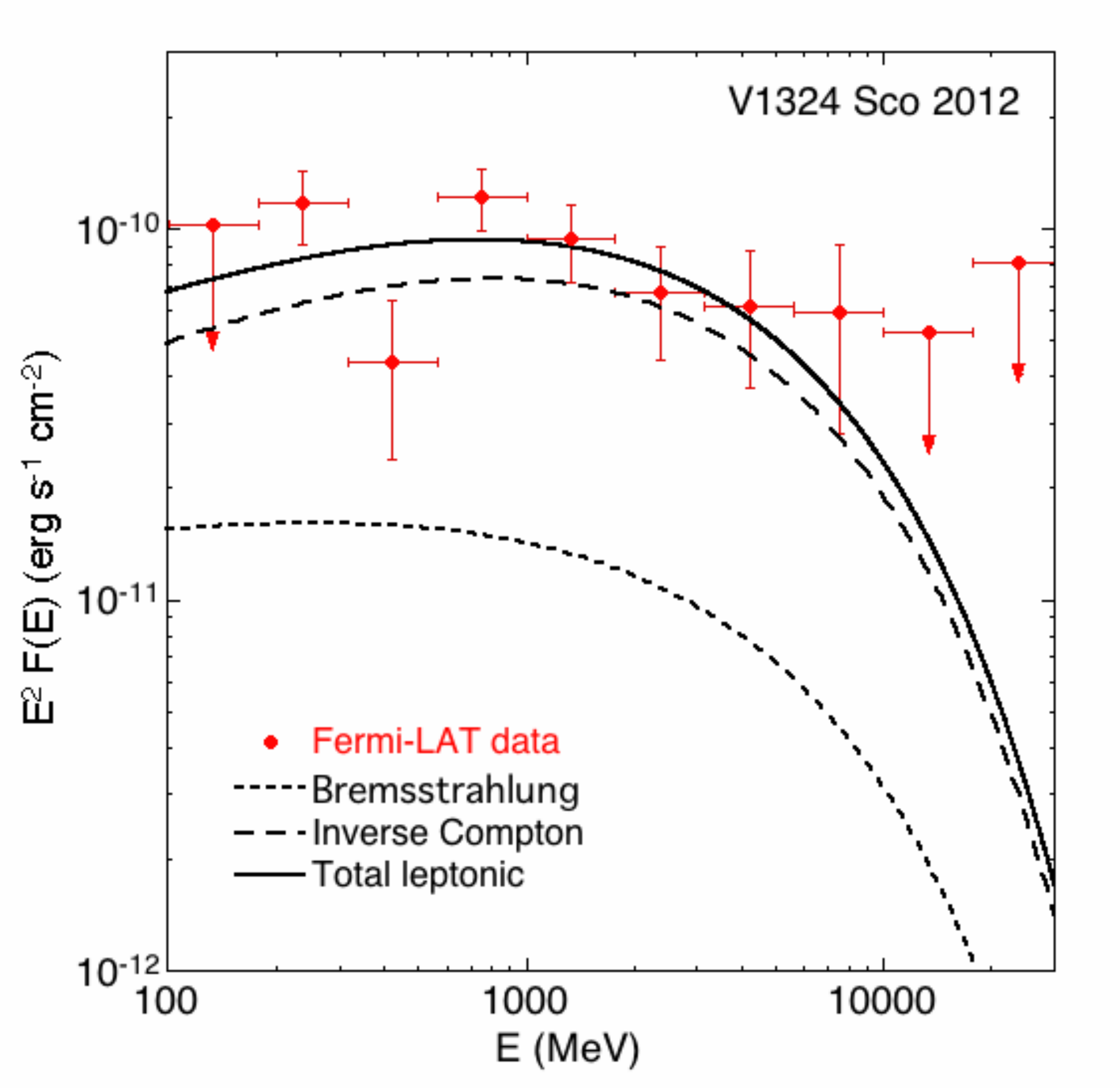}
    \includegraphics[width=7.75cm]{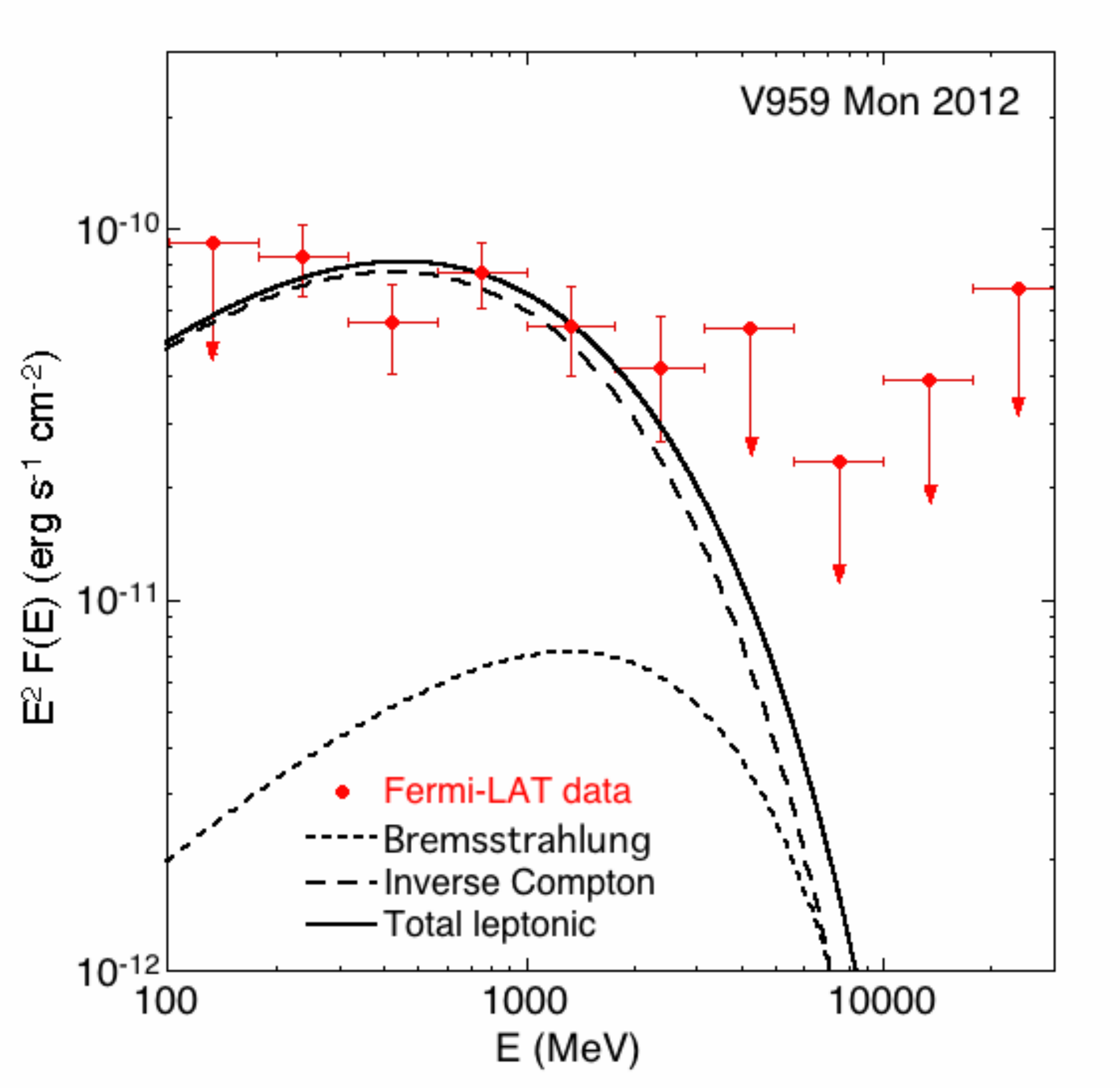}
    \includegraphics[width=7.75cm]{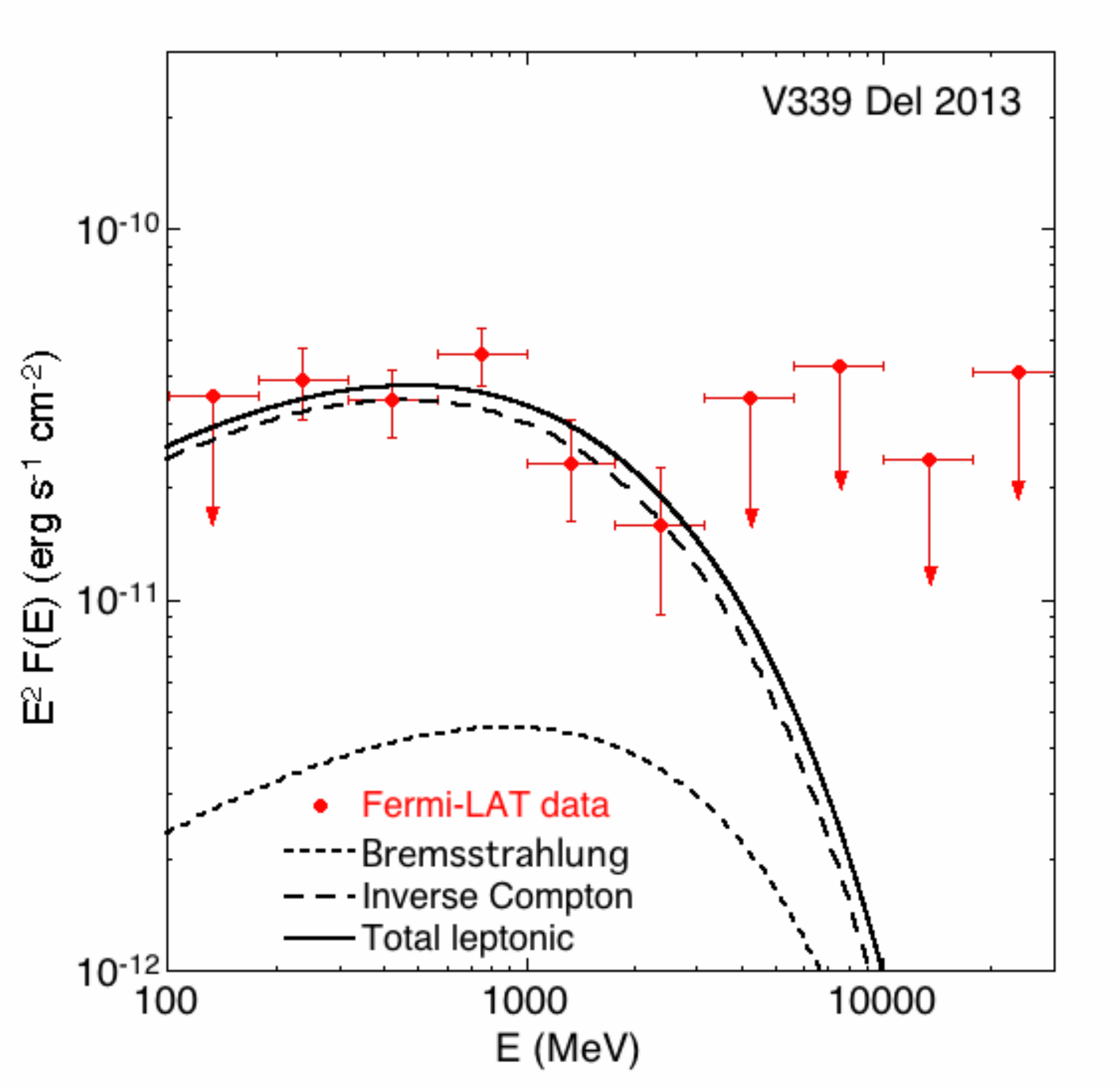}
  \end{center}
{\bf Fig.~S5.}
\Fermi-LAT $>$100 MeV average \gray\ spectra of the four novae over the full
17$-$27 day durations (as shown in Fig.~3) showing the inverse Compton and
bremsstrahlung spectral components in the leptonic model as well as their total.
\end{figure}

\newpage
\begin{figure}[htbp]
  \begin{center}
    \includegraphics[width=12.5cm]{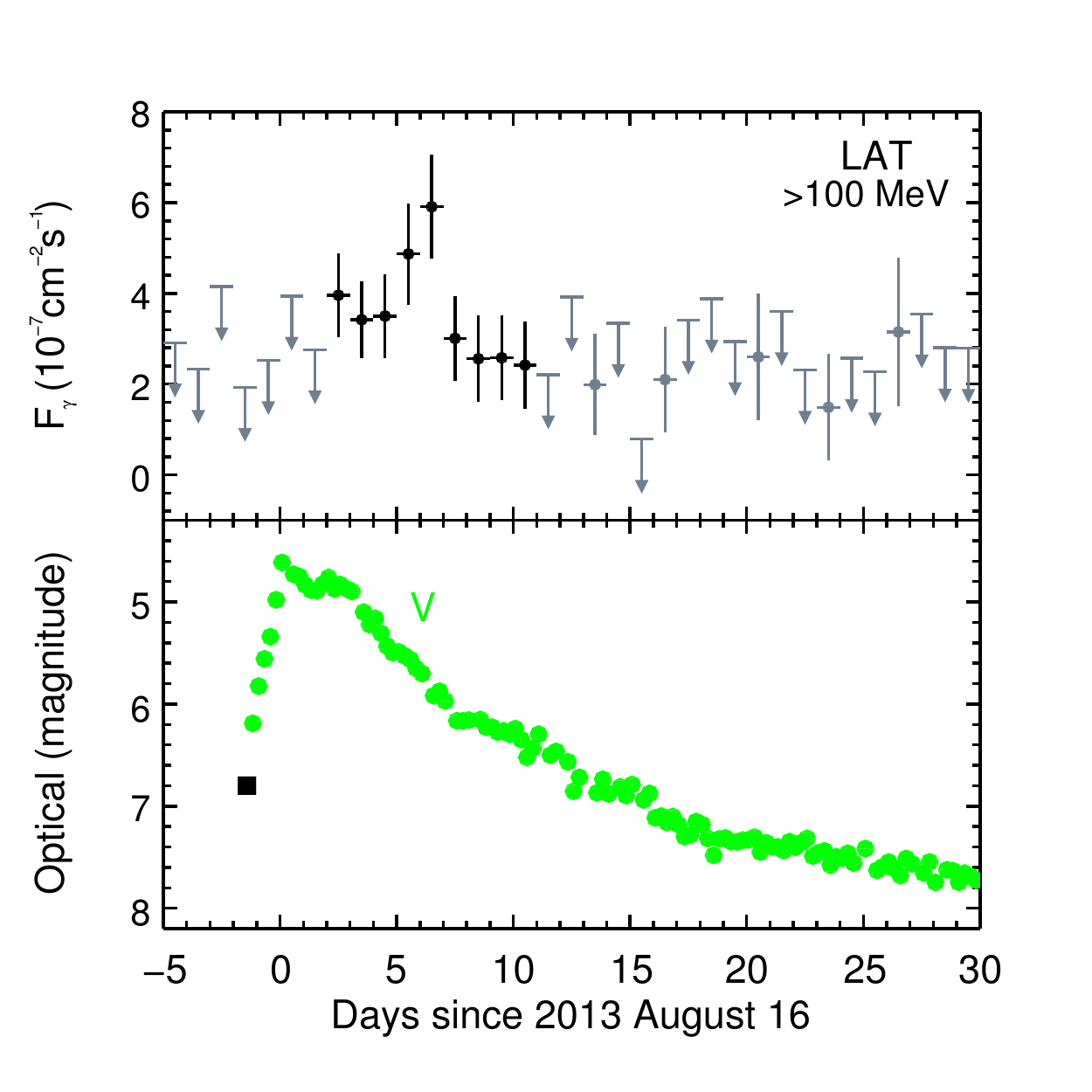}
  \end{center}
{\bf Fig.~S6.}
\Fermi-LAT 1-day binned light curve (top panel) of \ndel\ as shown in Fig.~2 
where vertical bars indicate 1$\sigma$ uncertainties for data points with $TS > 
9$ (black) and $TS=$ 4$-$9 (gray), while gray arrows indicate 2$\sigma$ limits 
when $TS<4$. The \gray\ light curve is compared to the 0.25 day binned optical 
$V$-band AAVSO light curve (green circles) for the same time interval (bottom 
panel) with the inclusion of the unfiltered discovery magnitude (black square) 
from \cite{ita13}.
\end{figure}


\newpage
\begin{sidewaystable*}
\small
  \begin{center}
    \tabcolsep 4.0pt
\begin{tabular}{lcccc}
\hline\hline
\multicolumn{1}{l}{Nova} &
\multicolumn{1}{c}{V407 Cyg 2010} &
\multicolumn{1}{c}{V1324 Sco 2012} &
\multicolumn{1}{c}{V959 Mon 2012} &
\multicolumn{1}{c}{V339 Del 2013} \\
\hline
\hline
\multicolumn{5}{c}{Single Power Law} \\
\hline
Flux            & $5.77\pm0.55$ & $5.90\pm0.87$ & $4.81\pm0.59$ & $2.26\pm0.28$ \\
Photon index    & $2.11\pm0.06$ & $2.16\pm0.09$ & $2.34\pm0.09$ & $2.26\pm0.08$ \\
TS              & 366.5         & 178.5         & 156.0         & 197.5 \\
\hline
\hline
\multicolumn{5}{c}{Exponentially Cutoff Power Law} \\
\hline
Flux            & $4.30\pm0.54$ & $4.93\pm2.45$ & $4.13\pm0.60$ & $1.93\pm0.29$ \\
Slope, $s$      & $1.23\pm0.20$ & $1.77\pm0.68$ & $1.69\pm0.25$ & $1.72\pm0.24$ \\
$E_{\rm c}$     & $1.44\pm0.36$ & $4.10\pm6.14$ & $1.45\pm0.61$ & $1.94\pm0.95$ \\
TS              & 408.1         & 184.6         & 170.3         & 208.9 \\
\hline
\multicolumn{5}{c}{Exponentially Cutoff Power Law split into intervals: (a), (b)} \\
\hline
MJD start       & 55265, 55271 & 56093, 56099 & 56097, 56103 & 56520, 56526 \\
Interval (days) & 6, 16        & 6, 11        & 6, 16        & 6, 21        \\
Flux            & $7.39\pm1.07$, $2.84\pm0.61$ & $6.11\pm1.37$, $3.92\pm1.10$ & $7.33\pm1.47$, $3.32\pm0.66$ & $2.81\pm0.54$, $1.56\pm0.34$ \\
Slope, $s$      & $1.39\pm0.22$, $1.00\pm0.38$ & $1.94\pm0.23$, $1.53\pm0.35$ & $1.35\pm0.38$, $1.92\pm0.31$ & $1.75\pm0.31$, $1.65\pm0.35$ \\
$E_{\rm c}$     & $1.78\pm0.63$, $1.14\pm0.43$ & $7.08\pm6.00$, $2.25\pm1.33$ & $0.95\pm0.43$, $1.97\pm1.42$ & $2.21\pm1.69$, $1.67\pm1.03$ \\
TS              & 261.7, 176.6 & 110.9, 76.7  & 117.3, 73.9  & 108.3, 108.3 \\
\hline
\hline
\multicolumn{5}{c}{Hadronic Modeling} \\
\hline
Flux           & $4.05^{+0.26}_{-0.37}$ & $4.25^{+0.54}_{-0.77}$ & $3.66^{+0.44}_{-0.46}$ & $1.75^{+0.23}_{-0.20}$ \\
$E_{\rm cp}$   & $10^{+1.0}_{-0.7}$ & $>32$ & $>3.2$ & $>10$ \\
Slope, $s_{\rm p}$ & $1.4^{+0.3}_{-0.4}$ & $2.6^{+0.2}_{-0.7}$ & $2.4^{+0.9}_{-1.4}$ & $2.6^{+0.5}_{-0.8}$ \\
TS             & 402.7 & 180.7 & 168.4 & 211.2 \\
\hline
\hline
\multicolumn{5}{c}{Leptonic Modeling} \\
\hline
Flux  		& $4.12^{+0.27}_{-0.50}$ & $4.91^{+0.64}_{-0.87}$ & $3.96^{+0.62}_{-0.55}$ & $1.93^{+0.16}_{-0.34}$ \\
$E_{\rm ce}$	& $1.78\pm0.05$ & $10^{+22}_{-7}$ & $1.8^{+3.8}_{-0.8}$ & $3.2^{+2.5}_{-1.4}$ \\
Slope, $s_{\rm e}$ & $-1.0^{+0.25}_{-0.5}$ & $2.0^{+0.5}_{-1.5}$ & $0.25^{+1.75}_{-0.55}$ & $1.25^{+0.75}_{-1.75}$ \\
TS  		& 403.1 & 181.8 & 169.7 & 210.0 \\
\hline
\hline
\end{tabular}
\normalsize
\label{table-gamma}
  \end{center} 
{\bf Table~S1.}
LAT \gray\ spectral fit results, model parameters, and best-fit results from 
the hadronic and leptonic modeling. Fluxes at $>$100 MeV energies are in units 
of $10^{-7}$ \phflux\ and the cutoff energies $E_{\rm c}$ in GeV. The reported 
errors on spectral parameters are 1$\sigma$ uncertainties and statistical only.  
We estimate that the systematic uncertainties are comparable or smaller, 
$\sim$8$\%$ for the fluxes and $\sim$0.1 in photon indices \cite{ack12s}.
\end{sidewaystable*} 

\newpage
\begin{table*}
\small
  \begin{center}
\begin{tabular}{lcccccccc}
\hline\hline
\multicolumn{1}{l}{Days since} &
\multicolumn{2}{c}{V407 Cyg 2010} &
\multicolumn{2}{c}{V1324 Sco 2012} &
\multicolumn{2}{c}{V959 Mon 2012} &
\multicolumn{2}{c}{V339 Del 2013} \\
\multicolumn{1}{l}{start} &
\multicolumn{1}{c}{$TS$} &
\multicolumn{1}{c}{$F_{\gamma}$} &
\multicolumn{1}{c}{$TS$} &
\multicolumn{1}{c}{$F_{\gamma}$} &
\multicolumn{1}{c}{$TS$} &
\multicolumn{1}{c}{$F_{\gamma}$} &
\multicolumn{1}{c}{$TS$} &
\multicolumn{1}{c}{$F_{\gamma}$} \\
\hline
\hline
  -4.5 &  3.0 & $<7.2$ &        0.0 & $<4.2$ &        0.3 & $<6.0$ &         0.0 & $<2.9$ \\ 	      
  -3.5 &  0.0 & $<3.0$ &        0.5 & $<4.2$ & 	      0.0 & $<4.8$ & 	     0.0 & $<2.3$ \\ 	      
  -2.5 &  0.0 & $<2.2$ &        0.2 & $<3.6$ & 	      0.5 & $<6.7$ & 	     1.7 & $<4.2$ \\ 	      
  -1.5 &  3.5 & $<6.6$ &        2.7 & $<6.1$ & 	      0.2 & $<5.8$ & 	     0.0 & $<1.9$ \\ 	      
  -0.5 &  0.0 & $<3.4$ &        3.0 & $<7.1$ & 	      1.7 & $<9.1$ & 	     0.0 & $<2.5$ \\ 	      
   0.5 & 16.2 & $6.2\pm2.2$ &   9.0 & $3.7\pm1.9$ &   9.0 & $5.4\pm2.5$ &    3.3 & $<3.9$ \\ 	      
   1.5 & 58.9 & $11.1\pm2.3$ & 14.1 & $6.9\pm2.3$ &   9.6 & $7.6\pm3.2$ &    1.1 & $<2.8$ \\ 	      
   2.5 & 12.8 & $5.9\pm2.1$ &  14.9 & $6.7\pm2.4$ &  14.4 & $8.4\pm2.9$ &   42.6 & $4.0\pm0.9$ \\      
   3.5 & 68.6 & $13.1\pm2.5$ & 16.0 & $7.1\pm2.4$ &  32.8 & $12.9\pm3.5$ &  37.2 & $3.4\pm0.8$ \\      
   4.5 & 56.8 & $13.9\pm2.6$ & 23.0 & $8.8\pm2.7$ &  27.9 & $10.6\pm3.1$ &  25.3 & $3.5\pm0.9$ \\      
   5.5 & 47.9 & $13.2\pm2.7$ & 33.4 & $10.9\pm2.7$ & 27.7 & $13.8\pm3.7$ &  38.8 & $4.9\pm1.1$ \\      
   6.5 & 27.1 & $7.5\pm2.1$ &  35.0 & $12.3\pm2.9$ & 15.7 & $9.8\pm3.2$ &   65.7 & $5.9\pm1.1$ \\      
   7.5 & 10.1 & $5.6\pm2.3$ &  14.3 & $7.6\pm2.6$ &  11.3 & $8.0\pm3.1$ &   25.8 & $3.0\pm0.9$ \\      
   8.5 & 11.0 & $4.7\pm1.9$ &   2.1 & $<7.9$ & 	     29.2 & $13.5\pm3.6$ &  14.4 & $2.6\pm0.9$ \\     
   9.5 & 35.1 & $10.3\pm2.5$ & 15.5 & $10.3\pm3.3$ &  3.3 & $<8.3$ & 	    17.5 & $2.6\pm0.9$ \\    
  10.5 & 12.5 & $3.9\pm1.8$ &   2.6 & $<9.0$ & 	      1.7 & $<7.2$ & 	    10.8 & $2.4\pm1.0$ \\    
  11.5 & 20.6 & $6.5\pm2.3$ &  10.3 & $6.7\pm3.0$ &  15.1 & $9.1\pm3.1$ &    0.9 & $<2.2$ \\ 	      
  12.5 &  1.1 & $<5.3$ &        0.0 & $<5.8$ & 	      2.8 & $<8.5$ & 	     2.5 & $<3.9$ \\ 	      
  13.5 & 11.7 & $6.8\pm2.6$ &   0.0 & $<5.4$ & 	      4.8 & $4.8\pm2.7$ &    5.0 & $2.0\pm1.1$ \\     
  14.5 & 16.5 & $6.5\pm2.3$ &   7.0 & $9.5\pm4.3$ &   2.3 & $<7.0$ & 	     1.3 & $<3.3$ \\ 	      
  15.5 & 14.8 & $7.2\pm2.5$ &   1.7 & $<7.2$ & 	      0.0 & $<3.6$ & 	     0.0 & $<0.8$ \\ 	      
  16.5 &  1.1 & $<6.2$ &        6.4 & $8.7\pm4.3$ &   1.9 & $<3.6$ & 	     6.0 & $2.1\pm1.2$ \\    
  17.5 &  8.0 & $4.5\pm2.1$ &   0.7 & $<10.6$ &       0.0 & $<1.6$ & 	     1.0 & $<3.4$ \\ 	      
  18.5 &  0.0 & $<3.0$ &        0.0 & $<6.5$ & 	      0.0 & $<2.4$ & 	     1.9 & $<3.9$ \\ 	      
  19.5 &  4.4 & $3.0\pm2.0$ &   0.9 & $<9.6$ & 	      6.3 & $2.6\pm1.3$ &    1.4 & $<2.9$ \\ 	      
  20.5 &  1.4 & $<7.5$ &        0.0 & $<31.1$ &       4.3 & $3.8\pm2.2$ &    7.9 & $2.6\pm1.4$ \\     
  21.5 &  4.1 & $3.8\pm2.3$ &   0.1 & $<7.6$ & 	      7.9 & $3.9\pm2.0$ &    0.4 & $<3.6$ \\ 	      
  22.5 &  2.3 & $<9.3$ &        0.0 & $<5.6$ & 	      0.0 & $<4.8$ & 	     0.0 & $<2.3$ \\ 	      
  23.5 &  0.1 & $<5.4$ &        2.5 & $<12.8$ &       0.3 & $<5.6$ & 	     4.1 & $1.5\pm1.2$ \\    
  24.5 &  0.0 & $<3.7$ &        0.0 & $<4.1$ & 	      0.0 & $<2.0$ & 	     0.1 & $<2.6$ \\ 	      
  25.5 &  3.0 & $<8.7$ &        0.4 & $<9.8$ & 	      0.0 & $<4.5$ & 	     0.0 & $<2.3$ \\ 	      
  26.5 &  0.5 & $<5.2$ &        0.4 & $<9.4$ & 	      2.5 & $<6.7$ & 	     6.1 & $3.1\pm1.6$ \\    
  27.5 &  3.4 & $<3.0$ &        0.0 & $<6.7$ & 	      0.0 & $<3.5$ & 	     0.1 & $<3.5$ \\ 	      
  28.5 &  0.0 & $<1.5$ &        0.0 & $<6.2$ & 	      0.6 & $<5.7$ & 	     0.0 & $<2.8$ \\ 	      
  29.5 &  0.3 & $<3.0$ &        0.3 & $<9.4$ & 	      2.9 & $<7.6$ & 	     0.0 & $<2.8$ \\         
\hline
\hline
\end{tabular}
\normalsize
\label{table-lc}
  \end{center} 
{\bf Table~S2.} 
Daily LAT \gray\ $TS$ and $>$100 MeV fluxes $F_{\gamma}$ in units of $10^{-7}$ 
\phflux\ and $95\%$ confidence upper limits (when $TS<4$) corresponding to 
Fig.~2 in the main paper. The dates indicated are the centers of the one day 
bins relative to the defined start times (\t0). 
\end{table*}

\newpage
\begin{table*}
\small
  \begin{center}
\begin{tabular}{lcccc}
\hline\hline
\multicolumn{1}{l}{Nova} &
\multicolumn{1}{c}{V407 Cyg 2010} &
\multicolumn{1}{c}{V1324 Sco 2012} &
\multicolumn{1}{c}{V959 Mon 2012} &
\multicolumn{1}{c}{V339 Del 2013} \\
\hline
\hline
$\dot{\epsilon}^{\pi^0}_{\gamma}$ (10$^{35}$ erg s$^{-1}$) & $3.04^{+0.04}_{-0.03}$ & $9.12^{+0.54}_{-0.93}$ & $3.50^{+0.27}_{-0.14}$ & $2.52^{+0.23}_{-0.11}$ \\
$M_{\rm ej}$ ($M_{\odot}$)  & 10$^{-6}$ & 10$^{-5}$ & 6$\times$10$^{-5}$ & 8$\times$10$^{-5}$ \\
$v_{\rm ej}$ (km s$^{-1}$)  & 3200 & 2200 & 3000 & 2000 \\
$\epsilon_{\rm p}$ (10$^{42}$ erg) & $6.8^{+0.4}_{-0.5}$ & $17.2^{+7.8}_{-1.5}$ & $10.1^{+5.4}_{-2.6}$ & $3.0^{+0.9}_{-0.7}$ \\
$\eta_{\rm p}$ ($\%$) & $6.6^{+0.4}_{-0.5}$ & $3.7^{+0.2}_{-0.4}$ & $0.19^{+0.10}_{-0.05}$ & $0.09^{+0.03}_{-0.02}$ \\
$F_{\epsilon}^{\pi^0}$ (10$^{-10}$ erg s$^{-1}$ cm$^{-2}$) & $3.46^{+0.04}_{-0.03}$ & $3.73^{+0.22}_{-0.38}$ & $2.23^{+0.17}_{-0.09}$ & $1.18^{+0.11}_{-0.05}$ \\
\hline
\hline
\end{tabular}
\normalsize
\label{table-modelpi0}
  \end{center} 
{\bf Table~S3.}
Best-fit \gray\ luminosity for the hadronic models 
($\dot{\epsilon}^{\pi^0}_{\gamma}$) and ejecta parameters used to calculate the 
energetics of the novae. $M_{\rm ej}$ is the mass of the ejecta, $v_{\rm ej}$ is 
the ejecta velocity, $\epsilon_{\rm p}$ is the total energy in protons, 
$\eta_{\rm p}$ is the conversion efficiency, and $F_{\epsilon}^{\pi^0}$ is the 
total energy flux. The 1$\sigma$ uncertainties were obtained from the fits of 
the models to the LAT data. 
\end{table*}

\begin{table*}
\small
  \begin{center}
\begin{tabular}{lcccc}
\hline\hline
\multicolumn{1}{l}{Nova} &
\multicolumn{1}{c}{V407 Cyg 2010} &
\multicolumn{1}{c}{V1324 Sco 2012} &
\multicolumn{1}{c}{V959 Mon 2012} &
\multicolumn{1}{c}{V339 Del 2013} \\
\hline
\hline
$\dot{\epsilon}^{\rm IC}_{\gamma}$ (10$^{35}$ erg s$^{-1}$) & 3.11$^{+0.04}_{-0.06}$  & 9.42$^{+1.04}_{-1.18}$ & 3.80$^{+0.36}_{-0.29}$ & 2.67$^{+0.15}_{-0.20}$ \\
$\epsilon_{\rm e}$ (10$^{41}$ erg) & $5.9\pm0.1$ & $13.4^{+3.1}_{-2.1}$ & $8.0^{+4.4}_{-1.1}$ & $6.6^{+1.3}_{-0.7}$ \\
$\eta_{\rm e}$ ($\%$) & $0.57\pm0.01$ & $0.28^{+0.06}_{-0.04}$ & $0.09^{+0.05}_{-0.01}$ & $0.17^{+0.03}_{-0.02}$ \\
$F_{\epsilon}^{\rm IC}$ (10$^{-10}$ erg s$^{-1}$ cm$^{-2}$) & 3.53$^{+0.05}_{-0.07}$  & 3.85$^{+0.43}_{-0.48}$ & 2.43$^{+0.23}_{-0.19}$ & 1.25$^{+0.07}_{-0.09}$ \\
\hline
\hline
\end{tabular}
\normalsize
\label{table-modelIC}
  \end{center} 
{\bf Table~S4.}
Best-fit \gray\ luminosity for the leptonic models ($\dot{\epsilon}^{\rm 
IC}_{\gamma}$), total energy in electrons ($\epsilon_{\rm e}$), conversion 
efficiency ($\eta_{\rm e}$), and $F_{\epsilon}^{\rm IC}$ the total energy flux 
obtained for the four novae. The 1$\sigma$ uncertainties were obtained from the 
fits of the models to the LAT data. 
\end{table*}

\newpage

\clearpage
\noindent {\bf The Fermi-LAT Collaboration:} 
M.~Ackermann$^{1}$, 
M.~Ajello$^{2}$, 
A.~Albert$^{3}$, 
L.~Baldini$^{4}$, 
J.~Ballet$^{5}$, 
G.~Barbiellini$^{6,7}$, 
D.~Bastieri$^{8,9}$, 
R.~Bellazzini$^{4}$, 
E.~Bissaldi$^{10}$, 
R.~D.~Blandford$^{3}$
% ...
\medskip
\begin{enumerate}
\item[1.] Deutsches Elektronen Synchrotron DESY, D-15738 Zeuthen, Germany
\item[2.] Department of Physics and Astronomy, Clemson University, Kinard Lab of Physics, Clemson, SC 29634-0978, USA
\item[3.] W. W. Hansen Experimental Physics Laboratory, Kavli Institute for Particle Astrophysics and Cosmology, Department of Physics and SLAC National Accelerator Laboratory, Stanford University, Stanford, CA 94305, USA
\item[4.] Istituto Nazionale di Fisica Nucleare, Sezione di Pisa, I-56127 Pisa, Italy
\item[5.] Laboratoire AIM, CEA-IRFU/CNRS/Universit\'e Paris Diderot, Service d'Astrophysique, CEA Saclay, 91191 Gif sur Yvette, France
\item[6.] Istituto Nazionale di Fisica Nucleare, Sezione di Trieste, I-34127 Trieste, Italy
\item[7.] Dipartimento di Fisica, Universit\`a di Trieste, I-34127 Trieste, Italy
\item[8.] Istituto Nazionale di Fisica Nucleare, Sezione di Padova, I-35131 Padova, Italy
\item[9.] Dipartimento di Fisica e Astronomia ``G. Galilei'', Universit\`a di Padova, I-35131 Padova, Italy
\item[10.] Istituto Nazionale di Fisica Nucleare, Sezione di Trieste, and Universit\`a di Trieste, I-34127 Trieste, Italy
\item[11.] NASA Goddard Space Flight Center, Greenbelt, MD 20771, USA
\item[12.] Laboratoire Univers et Particules de Montpellier, Universit\'e Montpellier 2, CNRS/IN2P3, Montpellier, France
\item[13.] Laboratoire Leprince-Ringuet, \'Ecole polytechnique, CNRS/IN2P3, Palaiseau, France
\item[14.] Consorzio Interuniversitario per la Fisica Spaziale (CIFS), I-10133 Torino, Italy
\item[15.] Istituto Nazionale di Fisica Nucleare, Sezione di Bari, 70126 Bari, Italy
\item[16.] INAF-Istituto di Astrofisica Spaziale e Fisica Cosmica, I-20133 Milano, Italy
\item[17.] Agenzia Spaziale Italiana (ASI) Science Data Center, I-00133 Roma, Italy
\item[18.] Center for Earth Observing and Space Research, College of Science, George Mason University, Fairfax, VA 22030, resident at Naval Research Laboratory, Washington, DC 20375, USA
\item[19.] Space Science Division, Naval Research Laboratory, Washington, DC 20375-5352, USA
\item[20.] Istituto Nazionale di Astrofisica - Osservatorio Astronomico di Roma, I-00040 Monte Porzio Catone (Roma), Italy
\item[21.] Department of Physics, Stockholm University, AlbaNova, SE-106 91 Stockholm, Sweden
\item[22.] The Oskar Klein Centre for Cosmoparticle Physics, AlbaNova, SE-106 91 Stockholm, Sweden
\item[23.] Royal Swedish Academy of Sciences Research Fellow, funded by a grant from the K. A. Wallenberg Foundation
\item[24.] The Royal Swedish Academy of Sciences, Box 50005, SE-104 05 Stockholm, Sweden
\item[25.] Institut universitaire de France, 75005 Paris, France
\item[26.] INAF Istituto di Radioastronomia, 40129 Bologna, Italy
\item[27.] Dipartimento di Astronomia, Universit\`a di Bologna, I-40127 Bologna, Italy
\item[28.] Dipartimento di Fisica, Universit\`a di Udine and Istituto Nazionale di Fisica Nucleare, Sezione di Trieste, Gruppo Collegato di Udine, I-33100 Udine
\item[29.] Universit\`a di Udine, I-33100 Udine, Italy
\item[30.] Dipartimento di Fisica ``M. Merlin" dell'Universit\`a e del Politecnico di Bari, I-70126 Bari, Italy
\item[31.] Center for Research and Exploration in Space Science and Technology (CRESST) and NASA Goddard Space Flight Center, Greenbelt, MD 20771, USA
\item[32.] Department of Physics and Department of Astronomy, University of Maryland, College Park, MD 20742, USA
\item[33.] Fermilab, Batavia, IL 60510, USA
\item[34.] Max-Planck-Institut f\"ur Radioastronomie, Auf dem H\"ugel 69, 53121 Bonn, Germany
\item[35.] Department of Physical Sciences, Hiroshima University, Higashi-Hiroshima, Hiroshima 739-8526, Japan
\item[36.] Istituto Nazionale di Fisica Nucleare, Sezione di Perugia, I-06123 Perugia, Italy
\item[37.] Dipartimento di Fisica, Universit\`a degli Studi di Perugia, I-06123 Perugia, Italy
\item[38.] NASA Postdoctoral Program Fellow, USA
\item[39.] Institut f\"ur Astro- und Teilchenphysik and Institut f\"ur Theoretische Physik, Leopold-Franzens-Universit\"at Innsbruck, A-6020 Innsbruck, Austria
\item[40.] Institute for Cosmic-Ray Research, University of Tokyo, 5-1-5 Kashiwanoha, Kashiwa, Chiba, 277-8582, Japan
\item[41.] Department of Physics and Center for Space Sciences and Technology, University of Maryland Baltimore County, Baltimore, MD 21250, USA
\item[42.] School of Physics and Astronomy, University of Southampton, Highfield, Southampton, SO17 1BJ, UK
\item[43.] Funded by a Marie Curie IOF, FP7/2007-2013 - Grant agreement no. 275861
\item[44.] Centre d'\'Etudes Nucl\'eaires de Bordeaux Gradignan, IN2P3/CNRS, Universit\'e Bordeaux 1, BP120, F-33175 Gradignan Cedex, France
\item[45.] CNRS, IRAP, F-31028 Toulouse cedex 4, France
\item[46.] GAHEC, Universit\'e de Toulouse, UPS-OMP, IRAP, Toulouse, France
\item[47.] Science Institute, University of Iceland, IS-107 Reykjavik, Iceland
\item[48.] CSIRO Astronomy and Space Science, Australia Telescope National Facility, Epping NSW 1710, Australia
\item[49.] Department of Astronomy, Stockholm University, SE-106 91 Stockholm, Sweden
\item[50.] Istituto Nazionale di Fisica Nucleare, Sezione di Torino, I-10125 Torino, Italy
\item[51.] Funded by contract ERC-StG-259391 from the European Community
\item[52.] Department of Astronomy, Department of Physics and Yale Center for Astronomy and Astrophysics, Yale University, New Haven, CT 06520-8120, USA
\item[53.] Department of Physics, Faculty of Science, Mahidol University, Bangkok 10400, Thailand
\item[54.] Hiroshima Astrophysical Science Center, Hiroshima University, Higashi-Hiroshima, Hiroshima 739-8526, Japan
\item[55.] Istituto Nazionale di Fisica Nucleare, Sezione di Roma ``Tor Vergata", I-00133 Roma, Italy
\item[56.] Center for Cosmology, Physics and Astronomy Department, University of California, Irvine, CA 92697-2575, USA
\item[57.] Department of Physics and Astronomy, University of Denver, Denver, CO 80208, USA
\item[58.] Max-Planck-Institut f\"ur Physik, D-80805 M\"unchen, Germany
\item[59.] Funded by contract FIRB-2012-RBFR12PM1F from the Italian Ministry of Education, University and Research (MIUR)
\item[60.] Department of Physics, University of Johannesburg, PO Box 524, Auckland Park 2006, South Africa
\item[61.] Santa Cruz Institute for Particle Physics, Department of Physics and Department of Astronomy and Astrophysics, University of California at Santa Cruz, Santa Cruz, CA 95064, USA
\item[62.] Department of Physics, The University of Hong Kong, Pokfulam Road, Hong Kong, China
\item[63.] National Research Council Research Associate, National Academy of Sciences, Washington, DC 20001, resident at Naval Research Laboratory, Washington, DC 20375, USA
\item[64.] NYCB Real-Time Computing Inc., Lattingtown, NY 11560-1025, USA
\item[65.] Institute of Space and Astronautical Science, Japan Aerospace Exploration Agency, 3-1-1 Yoshinodai, Chuo-ku, Sagamihara, Kanagawa 252-5210, Japan
\item[66.] Astronomical Observatory, Jagiellonian University, 30-244 Krak\'ow, Poland
\item[67.] Department of Chemistry and Physics, Purdue University Calumet, Hammond, IN 46323-2094, USA
\item[68.] Department of Physics, Graduate School of Science, Kyoto University, Kyoto, Japan
\item[69.] Institut de Ci\`encies de l'Espai (IEEE-CSIC), Campus UAB, 08193 Barcelona, Spain
\item[70.] Instituci\'o Catalana de Recerca i Estudis Avan\c{c}ats (ICREA), Barcelona, Spain
\item[71.] 3-34-1 Nishi-Ikebukuro,Toshima-ku, Tokyo 171-8501, Japan
\item[72.] Department of Physics, Center for Cosmology and Astro-Particle Physics, The Ohio State University, Columbus, OH 43210, USA
\item[73.] Praxis Inc., Alexandria, VA 22303, resident at Naval Research Laboratory, Washington, DC 20375, USA
\item[74.] Durtal Observatory, 6 Rue des Glycines, F-49430 Durtal, France
\item[75.] Dipartimento di Fisica ``Enrico Fermi", Universit\`a di Pisa, Pisa I-56127, Italy
\item[76.] Hamburger Sternwarte, Gojenbergsweg 112, 21029, Hamburg, Germany                      
\item[77.] Ammon, ID  83401, USA
\item[78.] INAF Osservatorio Astronomico di Trieste, Via G. B. Tiepolo 11, 34131 Trieste, Italy
\item[79.] American Astronomical Society, 2000 Florida Ave NW, Washington, DC 20009-1231, USA
\item[80.] School of Earth and Space Exploration, Arizona State University, PO Box 871404, Tempe, AZ 85287-1404, USA
\item[81.] 67 Rue Jacques Daviel, Rouen 76100, France
\end{enumerate}

\end{document}